\documentclass[useAMS,usenatbib]{mn2e}

 \usepackage{times}
 \usepackage{graphicx}

\def\pcm3{{\rm\thinspace cm$^{-3}$}}

\def\contcaption{\@conttrue\SFB@caption\@captype}

\title[A wide deep infrared look at the Pleiades with UKIDSS]
{A wide deep infrared look at the Pleiades with UKIDSS: 
new constraints on the substellar binary fraction and the low mass IMF
\thanks{Based on observations made with the United Kingdom Infrared
Telescope, operated by the Joint Astronomy Centre on behalf of the
U.K. Particle Physics and Astronomy Research Council.}}
\author[N. Lodieu et al.]{N. Lodieu$^{1,2}$\thanks{E-mail:
nlodieu@iac.es}, P. D. Dobbie$^{3,2}$, N. R. Deacon$^{4}$,
S. T. Hodgkin$^{5}$, N. C. Hambly$^{6}$,
 \newauthor
R. F. Jameson$^{2}$ \\
$^{1}$Instituto de Astrof\'isica de Canarias, V\'ia L\'actea s/n,
E-38205 La Laguna, Tenerife, Spain \\
$^{2}$Department of Physics and Astronomy, University of Leicester, 
University Road, Leicester LE1 7RH, UK \\
$^{3}$Anglo-Australian Observatory, PO Box 296, Epping, NSW, 1710, 
Australia \\
$^{4}$Department of Astrophysics, Radboud University Nijmegen, 
P.O. Box 9010, 6500 GL Nijmegen, The Netherlands \\
$^{5}$Institute of Astronomy, Madingley Road, Cambridge CB3 0HA \\
$^{6}$Scottish Universities' Physics Alliance (SUPA),
Institute for Astronomy, School of Physics, University of Edinburgh,
Royal Observatory, Blackford Hill, \\
Edinburgh EH9 3HJ, UK} 
\begin{document}

\date{Accepted \today{}. Received \today{}; in original form \today{}}

\pagerange{\pageref{firstpage}--\pageref{lastpage}} \pubyear{2002}

\maketitle

\label{firstpage}

\begin{abstract}
We present the results of a deep wide-field near-infrared survey
of 12 square degrees of the Pleiades conducted as part of the
UKIDSS Deep Infrared Sky Survey (UKIDSS) Galactic Cluster Survey (GCS).
We have extracted over 340 high probability proper motion members
down to 0.03 solar masses (M$_{\odot}$) using a combination of
UKIDSS photometry and proper motion measurements obtained by
cross-correlating the GCS with data from the Two Micron All Sky Survey 
(2MASS), the Isaac Newton (INT) and the Canada-France-Hawai'i (CFHT)
telescopes. Additionally, we have unearthed 73 new candidate
brown dwarf members on the basis of five band UKIDSS photometry alone.
We have identified 23 substellar multiple system candidates
out of 63 candidate brown dwarfs from the ($Y-K$,$Y$) and
($J-K$,$J$) colour-magnitude diagrams, yielding a binary
frequency of 28--44\% in the 0.075--0.030 M$_{\odot}$ mass range.
Our estimate is three times larger than the binary fractions
reported from high-resolution imaging surveys of field ultracool dwarfs 
and Pleiades brown dwarfs. However, it is marginally consistent with our
earlier ``peculiar'' photometric binary fraction of 50$\pm$10\% presented
in \citet{pinfield03}, in good agreement with the 32--45\% binary fraction 
derived from the recent Monte-Carlo simulations of \citet{maxted05} and 
compatible with the 26$\pm$10\% frequency recently estimated by 
\citet{basri06}. A tentative estimate of the mass ratios from 
photometry alone seems to support the hypothesis that binary brown 
dwarfs tend to reside in near equal-mass ratio systems.
In addition, the recovery of four Pleiades members targeted by
high-resolution imaging surveys for multiplicity studies suggests
that half of the binary candidates may have separations below the 
resolution limit of the {\it Hubble Space Telescope} or current 
adaptive optics facilities at the distance of the Pleiades 
(a $\sim$ 7 AU). 
Finally, we have derived luminosity and mass functions from the 
sample of photometric candidates with membership probabilities. 
The mass function is well modelled by a log-normal peaking at 
0.24 M$_{\odot}$ and is in agreement with previous studies in 
the Pleiades.
\end{abstract}


\section{Introduction}

Over the past two decades star forming regions and rich, young open 
clusters have been the focal points of numerous searches for substellar 
objects \citep[e.g.][]{jameson89,luhman99a,lucas00,bejar01,moraux03,lodieu06,lodieu07a}.
Part of the reasoning behind this is that in these environments 
brown dwarfs (BDs), which cool and fade after a brief period of 
deuterium burning \citep[$\sim$2--20 Myr;][]{baraffe98,palla05},
are comparatively luminous, allowing 2/4m class telescopes 
to probe to very low-masses. Furthermore, the members of such an 
aglomeration likely have a common age, distance and composition making 
a theoretical interpretation of their observed properties somewhat more 
straightforward (e.g.\ mass estimates).
The ultimate aim of these deep surveys is to characterise the substellar 
population, including the relative numbers of members as a function 
of mass, the lower mass limit to their manufacture, the binary 
fraction and the spatial distribution. Observational contraints 
on these properties can be used to critically examine
models of the star formation process.
Furthermore, comparison of these properties as derived in a number of 
different environments can provide clues as to whether the Initial 
Mass Function (IMF; dN/d$m$) and the binary fraction, as a function of mass,
are universal or dependent on the conditions in the nascent molecular 
clouds \citep{briceno02,luhman04a}.

The rich Pleiades cluster has been subjected to a particularly high 
degree of scrutiny as it has a number of highly attractive properties. 
For example, its constituents share a sizeable common proper motion 
\citep[$\mu_\alpha\cos\delta$ = 19.15 and $\mu_\delta$ = $-$45.72 mas/yr;]
[]{robichon99} so it is relatively straightforward to discriminate members
from the general field star population. 
The distance and the age of the cluster are well constrained.
Estimates of the former, which includes 
a meticulous astrometric analysis of the binary Atlas, 
concentrate around 134 pc with an uncertainty of 5 pc
\citep{johnson57,gatewood00,pinfield00,southworth05}. 
The latter is estimated
to be $125\pm8$ Myrs based on the location of the ``lithium boundary'' 
\citep*{rebolo92} as observed in the spectra of low-mass members 
\citep{stauffer98}.
Additionally, reddening along the line of sight to the cluster 
is generally low, E($B-V$) = 0.03 \citep*{Odell94}.

Recent work on very young clusters (age $<$ 10 Myrs) and star-formation 
regions e.g. $\sigma$ Ori \citep{bejar01}, the Trapezium Cluster
\citep{muench02,slesnick04}, IC348 \citep{luhman98,luhman00b,muench03}, 
Upper Sco \citep{lodieu07a} suggests that the IMF
continues slowly rising, down to about m=0.01M$_{\odot}$,
at least in these environments. However, mass estimates of
young BDs (age around 1 Myr or less) derived from
the direct comparison of their observed properties to
the predictions of theoretical models should be treated with 
caution because evolutionary calculations are not yet
coupled to detailed simulations of the collapse and accretion
phase of star formation \citep{baraffe02}.
Similarly, \citet{hillenbrand04} showed that models tend to 
underestimate masses by a few tens of percent in the
m = 1.0--0.3 M$_{\odot}$ mass range from an analysis of 
available dynamical mass measurements of pre-main-sequence stars.
While it would be imprudent to claim that the interpretation of 
observational data related to members of the Pleiades
is completely free of such uncertainties,
given the greater maturity of the cluster it seems realistic 
to assume that these uncertainties are significantly reduced. 

Previous CCD based studies of the Pleiades indicate that the present 
day mass function, across the stellar/substellar boundary and 
down to m$\sim$0.03M$_{\odot}$ \citep[as derived using the NextGen
and DUSTY models;][]{baraffe98,chabrier00c},
can be represented to first order by a slowly rising power 
law model, dN/dm $\propto$ m$^{-\alpha}$. For example, from their 
CFHT survey conducted at $R_{\rm CFHT}$ and $I_{\rm CFHT}$ and covering 
2.5 square degrees, \citet{bouvier98} identified 17 candidate
BDs (I$_{\rm C} \geq$ 17.8) and derived a power law index 
of $\alpha$ = 0.6$\pm$0.1\@. From their 1.1 square degree INT 
survey conducted at $I_{\rm RGO}$ and $Z_{\rm RGO}$, with 
follow-up work undertaken in the $K$-band, 
\citet{dobbie02b} unearthed 16 candidate substellar 
members and found a power law index of $\alpha$ = 0.8$\pm$0.2 
to be consistent with their data. \citet{moraux03}
extended the CFHT survey to an area of 6.4 square degrees 
(at $I_{\rm CFHT}$ and $Z_{\rm CFHT}$) 
and unearthed a total of 40 candidate BDs. They applied 
statistical arguments to account for non-members in their 
sample and derived a power law index of $\alpha$ = 0.6$\pm$0.1\@.

In arguably the most comprehensive deep study of the Pleiades 
to date, in terms of the selection criteria, \citet{jameson02}
assembled a sample of candidate substellar cluster members from 
four relatively recent CCD surveys, the International Time 
Project survey \citep{zapatero99b}, the CFHT survey 
\citep{bouvier98,moraux01}, the Burrell Schmidt 
survey \citep{pinfield00} and the INT survey \citep{dobbie02b}.
As candidates were selected at the very least on the basis of 
photometry in three passbands, contamination in the final sample 
of 49 likely BD members was estimated to be relatively 
low ($\sim$10\%). This sample was used to derive an IMF 
power law index of $\alpha$ = 0.41$\pm$0.08 over the
m = 0.3--0.035 M$_{\odot}$ \citep{jameson02}. Moreover, 
\citet{pinfield03} used these objects to infer a low-mass
stellar/substellar binary fraction of 50$\pm$10\% in the Pleiades.
This estimate is at least twice as large as determinations based on
high-resolution imaging studies of both old field ultracool dwarfs 
\citep[10--20\%; e.g.][]{burgasser03a,close03,bouy03} and Pleiades 
BDs \citep[13.3$^{+13}_{-4}$\%][]{martin00a,martin03,bouy06a}. 
Additionally, a number of the most detailed current theoretical models 
of brown dwarf formation predict a substellar binary fraction of only 
$\sim$5\% due to the disruptive influence of dynamical interactions during 
the earliest stages of their formation \citep{bate02,delgado_donate03}.

However, more recently, \citet{maxted05} have conducted Monte-Carlo
simulations to reproduce the binary properties of low-mass stars and 
BDs extracted from a number of radial velocity surveys 
\citep{guenther03,kenyon05,joergens06a} and have argued that the 
overall binary frequency is more likely to be in the range 32--45\%, 
comparable to that of K/M dwarfs. Furthermore, \citet{basri06} have 
concluded that spectroscopic binaries (i.e.\ close systems) can
account for $\sim$11\% over the 0--6 au separation range and should
be added to those resolved in the course of high-resolution imaging
despite some overlap in the separation ranges probed by both
samples. This conclusion was also reached by \citet{maxted05},
implying that current binary estimates might actually be about
twice larger. Nevertheless, the orbital separation 
and mass ratio distributions of low-mass binaries do appear to differ 
from those of their higher-mass G--M counterparts 
\citep{duquennoy91,fischer92}: the low mass separation distribution 
peaks around 4--8 au and three-quarter of systems have mass ratio 
larger than q = 0.8 \citep{burgasser07a}.

In this paper we present the results of our analyse of about 12 square 
degree survey of the Pleiades in $ZYJHK$\footnote{$ZYJHK$ are WFCAM 
filters. Filters from other studies are labelled in the text with the 
name of the observatory/survey to avoid confusion. However, WFCAM filters
are labelled as such in the figures.} and released as part of the 
UKIDSS Galactic Cluster Survey Data Release 1 (DR1) on 21 July 2006 
\citep{warren07a}. In Section \ref{Pleiades:surveys} we present the 
multi-epoch observations considered in this study to extract 
photometric and PM candidate members of the Pleiades, including 2MASS 
\citep{cutri03}, INT \citep{dobbie02b,jameson02}, CFHT \citep{moraux03}, 
and the UKIDSS GCS (second epoch) surveys. 
In Section \ref{Pleiades:new_cand} we describe the photometric 
selection of candidate cluster members from various colour-magnitude 
diagrams (CMDs) and estimate proper motions (PMs) from a probabilistic 
analysis. In Section \ref{Pleiades:status_old_cand} we review the list 
of previously published members recovered by our survey. 
In Section \ref{Pleiades:binary} we discuss the photometric binary 
frequency in the substellar regime and compare it with previous 
estimates in the Pleiades and for ultracool field dwarfs. 
In Section \ref{Pleiades:IMF} we derive the cluster luminosity and 
mass function and discuss the observed features over the mass range 
probed by the GCS\@. Finally, we summarise our work in Section 
\ref{Pleiades:summary}.

%
%
\section{Surveys and datasets}
\label{Pleiades:surveys}
%
%
%
%
\begin{figure*}
   \centering
   \includegraphics[width=\linewidth]{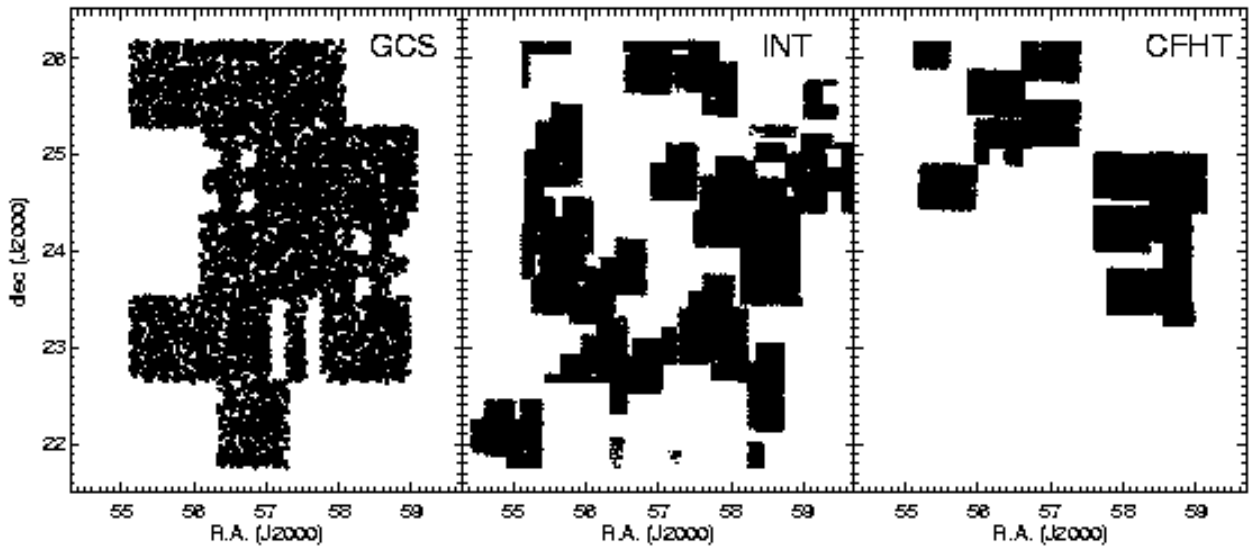}
   \caption{From left to right, coverage used in the Pleiades 
open cluster to derive PMs: coverage in $ZYJHK$ from the UKIDSS 
GCS DR1, INT optical survey \citep[e.g.][]{dobbie02a}, and CFHT 
$I_{\rm CFHT}$ and $Z_{\rm CFHT}$ study \citep{moraux03}. 
Only one in ten sources are plotted for the GCS coverage. 
The uneven coverage of the GCS and INT survey is due to the 
rejection of some tiles after quality control inspection.
}
   \label{fig_Pleiades:coverage}
\end{figure*}
\subsection{The UKIDSS GCS in the Pleiades}
\label{Pleiades:surveys_GCS}

The UKIRT Infrared Deep Sky Survey \citep{lawrence06}\footnote{The 
survey is described at www.ukidss.org}
is a deep large-scale infrared surveys conducted with
the UKIRT Wide-Field CAMera (WFCAM) on Mauna Kea in Hawai'i.
The survey is composed of 5 independent components: the Large Area Survey,
the Galactic Cluster Survey (hereafter GCS),
the Galactic Plane Survey, the Deep Extragalactic Survey,
and the Ultra-Deep Survey. All observations
are pipeline-processed at the Cambridge Astronomical Survey Unit
(CASU; Irwin et al.\ 2007, in preparation)\footnote{The CASU
WFCAM-dedicated  webpage can be found at http://apm15.ast.cam.ac.uk/wfcam}.
The processed data are then archived in Edinburgh and released to the
user community through the WFCAM Science Archive
(WSA; Hambly et al.\ 2007, in preparation)\footnote{The WFCAM
Science Archive is accessible at the URL http://surveys.roe.ac.uk/wsa}.
More details on the specificities of DR1 are given in \citet{warren07a}.

The WFCAM focal plane array consists of 4 Rockwell 2048\,$\times$\,2048
chips each covering a 13 arcmin by 13 arcmin field (or pawprint) with
a pixel scale of 0.4 arcsec. Each detector is separated by about
10 and 10 arcmin in the x and y axes, respectively. Consequently,
four pawprints are required to obtain a contiguous coverage
(or tile) of 0.8 square degrees (Casali et al.\ 2007, in preparation).
The $Z$ and $Y$ filters are centred at 0.92 and 1.03 $\mu$m, respectively,
and are 0.1 $\mu$m wide. The $J$, $H$, and $K$ near-infrared broad-band
filters are in the Mauna Kea Observatory (MKO) system \citep{hewett06}.

The GCS will cover $\sim$1000 square degrees in 10 star-forming
regions and open clusters down to $K$ = 18.4 mag at two epochs.
The main scientific driver of the GCS is to study the
IMF and its dependence with environment in the substellar regime
using an homogeneous set of observations of low-mass stars and BDs 
over a large area in several regions.
The UKIDSS DR1 contains 50 square degrees in the Pleiades,
Hyades, Taurus, and Orion \citep{warren07a}. The total area
released in the Pleiades in $ZYJHK$ is approximately
12 square degrees close to the central region of the cluster
(Figure \ref{fig_Pleiades:coverage}). The 100\% completeness
limits of the Pleiades GCS survey are  $Z \simeq$ 20.1, $Y \simeq$ 19.8,
$J \simeq$ 18.9, $H \simeq$ 18.4, and $K \simeq$ 17.8 mag.

\begin{figure}
   \includegraphics[width=1.00\linewidth]{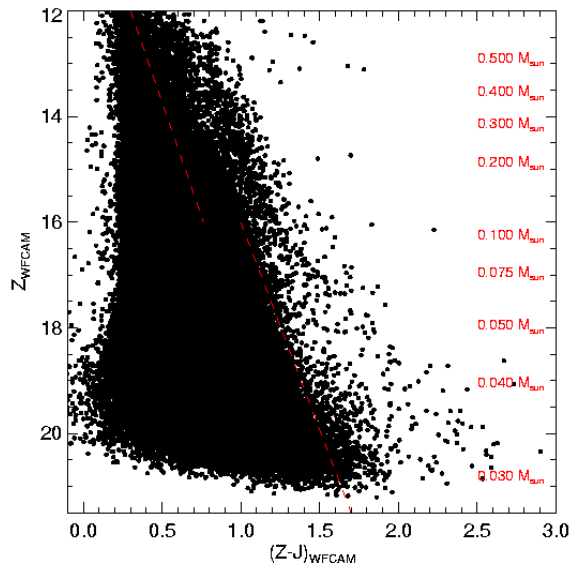}
   \caption{($Z-J$,$Z$) CMD for a 12
square degree area in the Pleiades cluster extracted from
the UKIDSS Galactic Cluster Survey Data Release 1\@.
The mass scale is shown on the right hand side of the diagrams
and extends down to 0.03 M$_{\odot}$, according to the
NextGen and DUSTY models \citep{baraffe98,chabrier00c}.
The photometric selection criteria applied to select new
candidates in the cluster prior to the derivation of proper
motions is shown as a red dashed line.
}
   \label{fig_Pleiades:ZJZcmd_alone}
\end{figure}

We have selected point sources in the Pleiades (UKIDSS GCS project \#2)
in a similar manner as described in our work conducted in Upper
Sco \citep[see SQL query in Appendix A of][]{lodieu07a}. The main
upgrades to the selection procedure applied in Upper Sco are two-fold.
Firstly, we insisted on detections classified as point sources
({\tt Class} parameters equal to $-$2 or $-$1) in all bands
and lifted the constraint on $Z$ and $Y$ by requesting detections
in the $JHK$ passbands only. Secondly, we have included a more 
relaxed constraint on the morphological shape
of the point sources in all passbands to increase the completeness
at the faint end of the survey i.e.\ we imposed that the
{\tt{ClassStat}} parameters lie between $-$3.0 and $+$3.0 in all
passbands unless undetected in the $Z$ and $Y$ filters.
The bright saturation limits are found to be $Z \simeq$ = 11.3, 
$Y \simeq$ 11.5, $J \simeq$ 11.0, $H \simeq$ 11.3, and 
$K \simeq$ 9.9 mag from the visual inspection of
the histogram of the number of stars as a function of magnitude.
However, we have only considered sources fainter than $Z$ = 12 mag
throughout this analysis to avoid saturated objects.
As for Upper Sco, the SQL query includes the cross-correlation
with 2MASS to compute PMs for all sources with a 2MASS counterpart
(the objects undetected in 2MASS are included in the catalogue
but PM is not available).
The query returned a total of 105,092 sources. The full coverage
is displayed in Fig.\ \ref{fig_Pleiades:coverage} and the resulting
($Z-J$,$Z$) CMD is shown in Fig.\ \ref{fig_Pleiades:ZJZcmd_alone}.
Note that the theoretical isochrones plotted in this paper were
specifically computed for the WFCAM set of filters and kindly
provided by Isabelle Baraffe and France Allard.

%
%
%

%
%
\begin{table}
  \caption{Main characteristics of the surveys (GCS, 2MASS, CFHT, 
and INT) used in this study to derive proper motions in the Pleiades 
cluster. We list the photometric passbands, the completeness
limit, the external astrometric accuracy (in arcsec), epoch of 
observations, and coverage (in square degrees) for each survey.
Note that for the CFHT survey, the 90\% completeness limit is quoted.
}
  \label{tab_Pleiades:surveys_char}
  \begin{tabular}{l c c c c c c c}
  \hline
Survey & Filters & Limit & Astrom.\   & Epoch &  Cov.\   \cr
       &         &  mag  &  arcsec    &  year &  deg$^{2}$\ \cr
  \hline
GCS   & $ZYJHK$         & $J \sim$ 18.6 & 0.1  & 2005--2006 &  12  \cr
2MASS & $JHK_{s}$       & $J \sim$ 15.8 & 0.1  & 1998--2000 & 300  \cr
CFHT  & $IZ_{\rm CFHT}$ & $Z \sim$ 22.0 & 0.2  & Dec 2000   & 6.4  \cr
INT   & $Z_{\rm RGO}$   & $Z \sim$ 18.5 & 0.3  & 1998--2000 &  20  \cr
 \hline
\end{tabular}
\end{table}

%
%
\subsection{The 2MASS survey in the Pleiades}
\label{Pleiades:surveys_2MASS}

As mentioned in the previous section, the SQL query includes a
cross-correlation of GCS sources with the nearest 2MASS counterpart
when available. The typical accuracy of the resulting proper motion 
measurement is less than 12 mas/yr down to the 2MASS 5$\sigma$ 
completeness limit of $J_{\rm 2MASS}$ = 15.8 mag for the 
5--7 year baseline. The coverage considered here overlaps 
with the study by \citet{adams01a} which covered the entire cluster 
up to 10 degrees away from the centre and was complete down to 
0.1 M$_{\odot}$ (Table \ref{tab_Pleiades:surveys_char}. 
The level of contamination of 
this earlier study was estimated to be less than 13\% down to 
$K_{\rm 2MASS}$ = 14 mag based on an extensive spectroscopic follow-up
program of several hundreds of photometric and PM candidates
drawn from a cross-correlation between 2MASS \citep{cutri03}
and the POSS I and POSS II plates \citep{reid91}.
However, as a result of less accurate PMs at fainter magnitudes 
($K_{\rm 2MASS}$ = 14--14.3 mag), contamination by field stars was at a 
much higher level, rendering membership assessement at the bottom of this
survey considerably less reliable.

%
%
\subsection{The INT-PL-IZ survey}
\label{Pleiades:surveys_INT}

Approximately 20 square degrees of the Pleiades was surveyed with the
INT and Wide Field Camera (WFC) over the course of several semesters 
between 1998-2000\@. The WFC consists of four 2048x4196
pixel EEV CCDs with each pixel corresponding to 0.33 arcsec on the sky. 
Data were obtained using the $Z_{\rm RGO}$ filter and exposure times 
ranging from 600--1200 seconds in typical seeing of 1.0--1.5 arcsec.
The data were reduced at the Cambridge Astronomical Survey Unit using the
WFC data reduction pipeline, details of which are given in \citet{irwin01}.
Morphological classification and aperture photometry was performed 
on all sources detected at a significance level of 7$\sigma$ or greater.
Coordinates are accurate to 0.3 arcsec externally
(Table \ref{tab_Pleiades:surveys_char}).
Plots of log10[number of sources] against magnitude indicate that 
the vast majority of the $Z_{\rm RGO}$-band images are photometrically 
complete to $Z_{\rm RGO} \le$ 18.5 mag, with some data complete to 
1.5 magnitudes deeper \citep[e.g.][]{dobbie02b} because about half of
the INT images were taken under non-photometric conditions. 
The overlap of the INT data with the GCS survey is depicted in 
the middle panel of Fig.\ \ref{fig_Pleiades:coverage}.

%
%
\subsection{The CFHT survey}
\label{Pleiades:surveys_CFHT}

The raw CFH12K data were extracted from the Canadian Astrophysical 
Data Center archive and were processed at Cambridge University using 
the same general purpose pipeline described in the previous section
\citep{irwin01}. Characteristics of the re-processed data are
given in Table \ref{tab_Pleiades:surveys_char} and the coverage
falling in the area covered by the GCS is shown in the right 
panel of Fig.\ \ref{fig_Pleiades:coverage}. The CFHT survey
(and the INT) were used, in this paper, to derive proper motions 
and not for photometric purposes.

%
%
%
\begin{figure*}
   \centering
   \includegraphics[width=\linewidth]{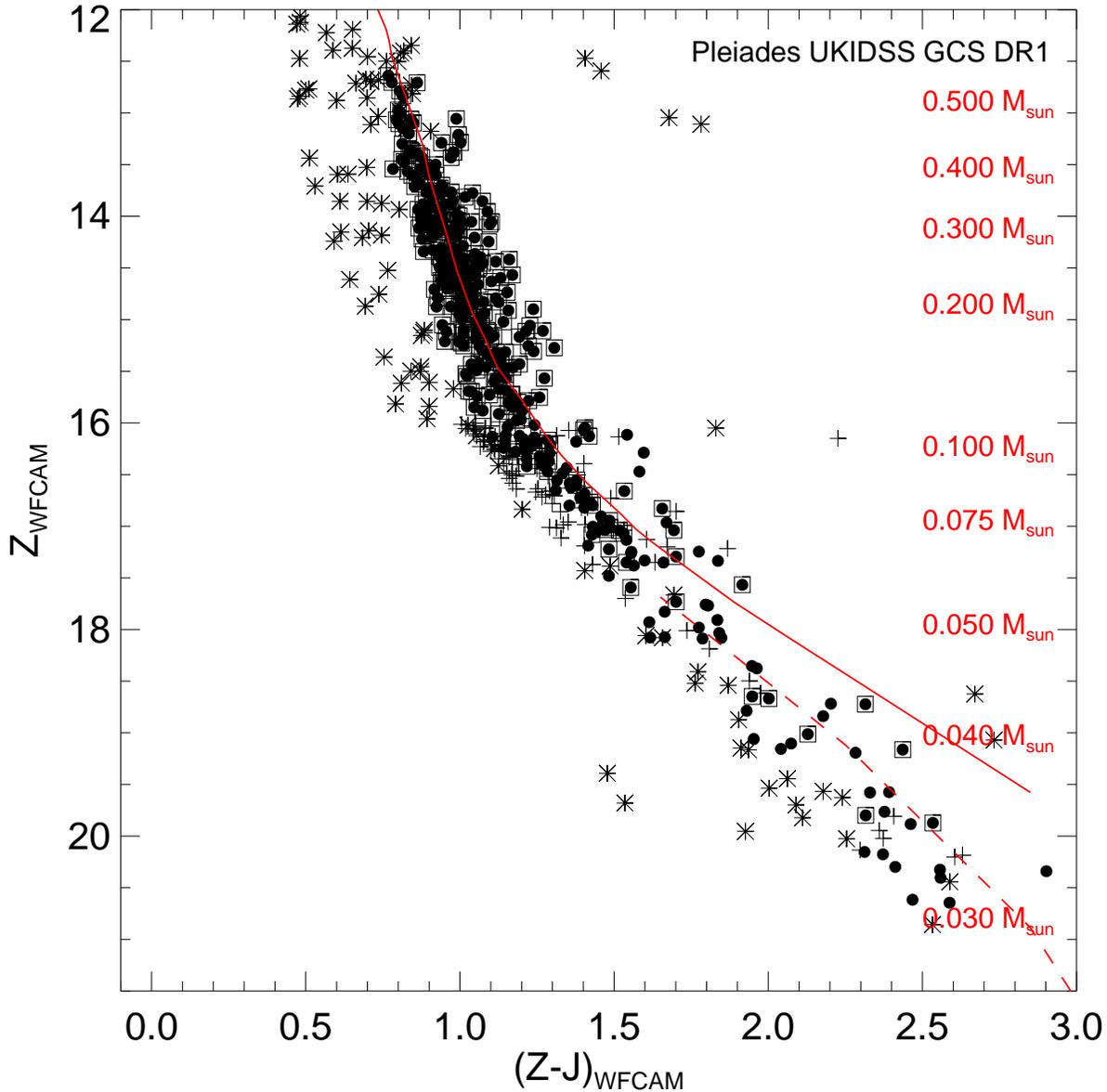}
   \caption{($Z-J$,$Z$) colour-magnitude diagram for all
   high-probability PM members in the Pleiades (filled dots
   with open squares) from
   the cross-correlation of the GCS DR1 with the 2MASS, INT, and
   CFHT surveys of the clusters. New photometric candidates are
   shown as filled dots whereas photometric and PM
   non-members are displayed as star symbols and crosses, respectively.
   Overplotted are the 120 Myr NextGen \citep[solid line;][]{baraffe98}
   and DUSTY \citep[dashed line;][]{chabrier00c} isochrones.
   The mass scale is shown on the right hand side of the diagrams
   and spans 0.5--0.03 M$_{\odot}$, according to the NextGen
   and DUSTY 120 Myr isochrones at the distance of the Pleiades.
}
   \label{fig_Pleiades:ZJZcmd}
\end{figure*}

%
%
\section{New substellar members in the Pleiades}
\label{Pleiades:new_cand}
\subsection{An outline of our selection method}

In this section we outline our method of selecting very-low-mass 
stellar and substellar candidate members of the Pleiades using five 
passband photometry and PMs. The main steps of the procedure are
as follows:
\begin{enumerate}
\item Make a conservative cut in the ($Z-J$,$Z$) CMD to select bright
($Z \leq$ 16 mag) and faint ($Z \geq$ 16 mag) photometric candidates 
(dashed lines in Fig.\ \ref{fig_Pleiades:ZJZcmd_alone}).
\item Obtain PM from the 2MASS vs GCS cross-correlation for sources
brighter than the 2MASS completeness at $Z \sim$ 17 mag and from 
the (INT$+$CFHT) vs GCS cross-correlation for objects fainter 
than $Z$ = 16 mag (Section \ref{Pleiades:new_cand_PM}).
\item Analyse the vector point diagram in a probabilistic manner
and infer a membership probability for each photometric candidate 
selected in the ($Z-J$,$Z$) CMD for which a proper motion measurement
exists (Section \ref{Pleiades:new_cand_proba}). Derive a list of probable 
cluster members by choosing a specific threshold for the membership
probability (p$\ge$0.6)
%
%
\item Weed out any remaining proper motion candidates that have
colours in the ($Y-J$,$Y$) and ($J-K$,$J$) CMDs which are inconsistent 
with cluster membership (Section \ref{Pleiades:new_cand_proba};
Table \ref{tab_Pleiades:ZJcand}).
\item Select new photometric candidates in the ($Z-J$,$Z$) CMD
for which no PM is available. Their membership will be constrained
further by examining their location in other CMDs and comparing it
to proper motion members (Section \ref{Pleiades:new_cand_photONLY})
\item Identify new low-mass BD candidates from the ($Y-J$,$Y$) CMD
i.e.\ sources undetected in the $Z$-band
\item Identify new possible low-mass BDs from the ($J-K$,$J$) CMD
i.e.\ sources undetected in the $Z$ and $Y$ passbands
\end{enumerate}

%
%
\subsection{Computation of proper motions}
\label{Pleiades:new_cand_PM}
%
%
\subsubsection{2MASS vs GCS cross-correlation}
\label{Pleiades:new_cand_2MASS}

As described in Section \ref{Pleiades:surveys_2MASS}, the SQL
query used to retrieve the full Pleiades catalogue from the
WSA includes a cross-correlation between the GCS and 2MASS source 
coordinates to derive PMs. Consequently, PMs are available for all sources 
brighter than the 2MASS 5$\sigma$ completeness limit 
($J_{\rm 2MASS}$ = 15.8 mag; $Z \sim$ 17.5 mag) with an accuracy 
better than
12 mas/yr. The vector point diagram of sources brighter than
$Z$ = 16 mag and located to the right of a line running from
($Z-J$,$Z$) = (0.3,12.0) to (1.4,21.5) is shown in the left
panel of Fig.\ \ref{fig_Pleiades:VPD}. The location of the
Pleiades cluster stands out clearly in this diagram due to its
large PM ($\mu_{\alpha}cos\delta \sim$20; $\mu_{\delta} \sim -$40
mas/yr). All sources in this diagram are associated with a 
membership probability and this sample is referred throughout
the remainder of this paper as the ``bright'' sample.

\subsubsection{INT and CFHT vs GCS cross-correlation}
\label{Pleiades:new_cand_INT_CFHT}

\begin{figure*}
   \centering
   \includegraphics[width=0.49\linewidth]{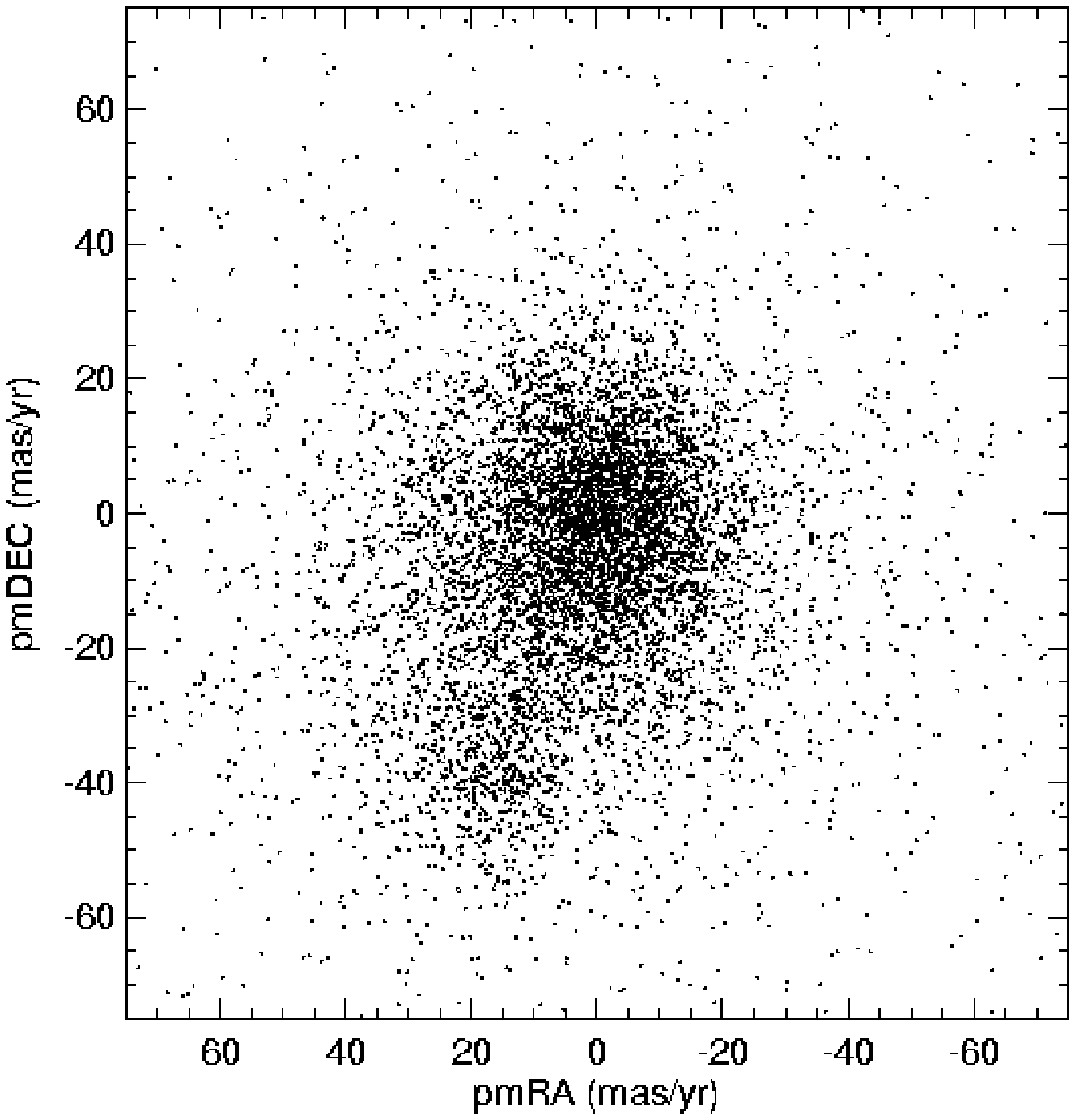}
   \includegraphics[width=0.49\linewidth]{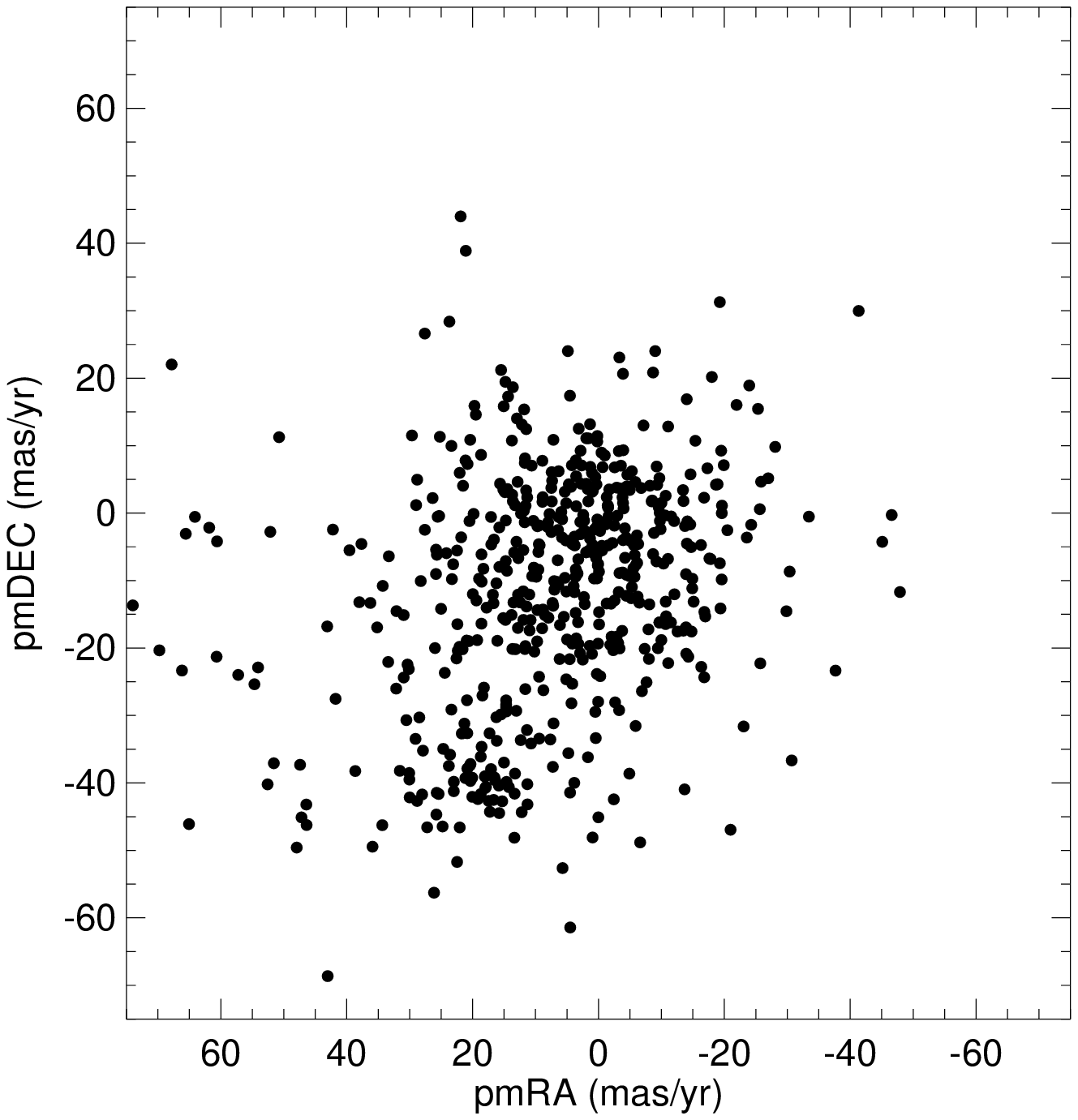}
   \caption{Vector point diagrams (PM in right ascension
   versus PM in declination) for candidates selected
   to the right of the dashed lines plotted in
   Fig.\ \ref{fig_Pleiades:ZJZcmd_alone}: left panel shows
   bright candidates ($Z \leq$ 16 mag) with 2MASS counterparts
   and the right panel displays fainter sources with INT and CFHT
   counterparts. The location of the Pleiades
   cluster is clearly seen around $+$20 and $-$40 mas/yr in right
   ascension and declination, respectively.
}
   \label{fig_Pleiades:VPD}
\end{figure*}

The completeness of the 2MASS survey is insufficient to derive PMs
over the full magnitude range probed by the GCS\@. However, the
Pleiades has been extensively surveyed in the optical over wide area
and deeply. Thus, we have cross-correlated the INT and CFHT
catalogues (described in Sections \ref{Pleiades:surveys_INT} 
and \ref{Pleiades:surveys_CFHT}) with the GCS to infer PMs
for sources fainter than $Z$ = 16 mag (hereafter ``faint'' sample).
Previous optical surveys have achieved similar depths to the GCS,
allowing the derivation of PMs down to $Z \sim$ 21 mag,
corresponding to masses of 30 M$_{\rm Jup}$ for cluster members
according to theoretical isochrones \citep{chabrier00c}.
We have adopted on purpose an overlapping range between the
``bright'' and ``faint'' samples to scale both luminosity
functions (see Section \ref{Pleiades:IMF}). The vector point 
diagram of GCS sources with optical counterparts is displayed
in the right panel of Fig.\ \ref{fig_Pleiades:VPD}.

The photometric catalogues of the GCS and the CFHT and the 
GCS and the INT surveys were cross correlated using a matching 
radius of 2''. To determine proper motions we constructed 12 
coefficient transforms between the coordinates of all sources 
common to both epochs using routines in the SLALIB package.
To minimise the effects of large scale astrometric distortions 
in the transforms this process was undertaken on a chip by chip 
basis. Despite the greater depth of the CFHT data relative to the 
INT data, the resulting proper motion uncertainties are very similar,
irrespective of epoch 1 dataset. These are typically found to be 7 
mas/yr at $Z$ = 15--18 mag but grow to 12 mas/yr at $Z$ = 19--20 mag.
  
The ($Z-J$,$Z$) CMD is shown in Fig.\ \ref{fig_Pleiades:ZJZcmd} 
whereas the corresponding vector point diagram is displayed 
in Fig.\ \ref{fig_Pleiades:VPD}.

%
%
\subsection{Membership probabilities}
\label{Pleiades:new_cand_proba}

In order to calculate formal membership probabilities we have used
the same technique as \citet{deacon04} to fit distribution functions
to proper motion vector point diagrams \citep{hambly95}. 
First we have rotated the vector point diagram so the cluster lies 
on the y-axis using the rotation transformation below 
(Equations \ref{eq:rot1} and \ref{eq:rot2}):
\begin{equation}
\mu_{x1} = 0.3896 \times \mu_{x} - 0.921 \times \mu_{y}
\label{eq:rot1}
\end{equation}
\begin{equation}
\mu_{y1} = 0.3896 \times \mu_{y} - 0.921 \times \mu_{x}
\label{eq:rot2}
\end{equation}
corresponding a rotation angle of 23.7 degrees, assuming a 
PM of (19.7,$-$44.82) mas/yr for the Pleiades.
 
We have assumed that there are two contibutions to the total 
distribution $\phi(\mu_{x},\mu_{y})$, one from the cluster 
($\phi_{c}(\mu_{x},\mu_{y})$) and one from the field stars
($\phi_{f}(\mu_{x},\mu_{y})$). The fitting region was delineated
by $-$50 $< \mu_{x} <$ 50 mas/yr and 20 $< \mu_{y} <$ 70 mas/yr.
These were added by means of a field star fraction $f$ to 
yield an expression for $\phi$ given in Equation \ref{eq:phi}:
\begin{equation}
\phi(\mu_{x},\mu_{y}) = f \phi_{f}(\mu_{x},\mu_{y}) + (1-f) \phi_{c}(\mu_{x},\mu_{y})
\label{eq:phi}
\end{equation}

We have assumed that the cluster distribution is characterised by
a two variable gaussian with a single standard deviation 
$\sigma$ and mean proper motion values in each axis 
$\mu_{xc}$ and $\mu_{yc}$ (Equation \ref{eq:phic}):

\begin{equation}
\phi_{c} \propto exp \left(-\frac{(\mu_{x} - \mu_{xc})^{2}+(\mu_{y} - \mu_{yc})^{2}}{2 \sigma^{2}}\right) 
\label{eq:phic}
\end{equation}

The field star distribution was fitted by a single gaussian in
the $x$ axis (with standard deviation $\Sigma_{x}$ and mean 
$\mu_{xf}$) and a declining exponential in the $y$ axis with 
a scale length $\tau$ (Equation \ref{eq:phif}).
The use of a declining exponential is a standard method
\citep[e.g.][]{jones91} and is justified in that the field star 
distribution is not simply a circularly-symmetric error 
distribution (i.e.\ capable of being modelled as a 2d 
Gaussian) - rather there is a prefered direction of
real field star motions resulting in a characteristic velocity
ellipsoidal signature, i.e.\ a non-Gaussian tail, in the 
vector point diagram. This is best modelled (away from the 
central error-dominated distribution) as an exponential in the 
direction of the antapex (of the solar motion).

\begin{equation}
\phi_{f} \propto exp \left(-\frac{(\mu_{x} - \mu_{xf})^{2}}{2 \Sigma_{x}^{2}} - \frac{\mu_{y}}{\tau}\right)
\label{eq:phif}
\end{equation}

Then, we have solved those equations for these seven parameters 
($f$, $\sigma$, $\mu_{xc}$, $\mu_{yc}$, $\Sigma_{x}$, $\mu_{xf}$, $\tau$).
This fitting process was tested by \citet{deacon04} where
simulated data sets were created and run through the fitting process
to recover the input parameters. These tests produced no significant 
offsets in the parameter values \citep[see Table 3 and Appendix A 
of][for results and more details on the procedure]{deacon04}.
Hence, we have calculated the formal membership probabilities as,

\begin{equation}
p=\frac{\phi_{c}}{f\phi_{f}+(1-f)\phi_{c}}
\end{equation}

%
%
%

%
%
\begin{table}
  \caption{Summary of the results after running the programme
to derive membership probabilities. For each $Z$ magnitude range,
we list the number of stars used in the fit (Nb), the field star 
fraction f, and parameters describing the cluster and field star 
distribution. Units are in mas/yr except for the number of 
stars and the field star fraction f. The cluster star distribution
is described by the mean proper motions in the x and y 
directions ($\mu_{x_{c}}$ and $\mu_{y_{c}}$) and a standard
deviation $\sigma$. Similarly, the field star distribution is
characterised by a scale length for the y axis ($\tau$), a
standard deviation $\Sigma_{x}$, and a mean proper motion
in the x direction ($\mu_{x_{f}}$). 
}
  \label{tab_Pleiades:prob_results}
  \begin{tabular}{@{\hspace{0mm}}c c c c c c c c c@{\hspace{0mm}}}
  \hline
$Z$ & Nb & f & $\sigma$ & $\mu_{x_{c}}$ & $\mu_{y_{c}}$ & $\tau$ & $\Sigma_{x}$& $\mu_{x_{f}}$ \cr
  \hline
10--12 &  12 & 0.55 &  7.66 & 2.38 & 43.91 &  6.18 & 13.58 &  5.10 \cr
12--14 & 300 & 0.61 &  5.95 & 0.48 & 40.34 & 17.79 & 20.47 &  2.55 \cr
14--16 & 616 & 0.63 &  6.65 & 0.20 & 41.71 & 16.55 & 17.02 &  4.21 \cr
16--18 &  82 & 0.53 &  4.36 & 1.81 & 42.97 & 22.45 & 21.96 &  6.81 \cr
18--19 &  16 & 0.68 &  6.53 & 1.80 & 42.98 & 26.15 & 12.91 &  7.83 \cr
19--20 &  29 & 0.68 &  8.05 & 1.80 & 42.98 & 14.76 & 11.45 & 12.07 \cr
$>$ 20 &  18 & 0.86 & 10.63 & 1.80 & 42.98 & 11.10 & 13.61 &  0.38 \cr
 \hline
\end{tabular}
\end{table}

As the astrometric errors varied in magnitude we have splitted the
sample into six bands. The first three bands (from $Z=12.0$
to $Z=18.0$) were each two magnitudes wide and were fitted
with all seven parameters in the same way as described in
\citet{deacon04}. As the astrometric errors increased rapidly
at the faint end, ranges only one magnitude wide were used.
In these bands the number of cluster stars was so small so that
we have fixed the location of the cluster on the vector point diagram
($\mu_{xc}$ and $\mu_{yc}$) to the values from a brighter bin.
The other parameters were fitted as normal. A summary of the
fitted parameters from the probabilistic analysis described
above is given in Table \ref{tab_Pleiades:prob_results}.

This sample of 1061 sources with membership probabilities
(hereafter ``PM sample'') is
used to derive the luminosity and mass function in
Section \ref{Pleiades:IMF}. Among them, 379 have probabilities
equal or higher than 0.6 (or 60\%; named hereafter as ``high
probability PM members''), including 75 that we have
classified as photometric non-member after examining their position
in several CMDs. The mean probability for high probability 
members is 0.775\@. In Table \ref{tab_Pleiades:ZJcand}, we list
all high probability PM members fainter than $Z$ = 16 mag
i.e.\ masses below 0.1 M$_{\odot}$ according to the NextGen
models \citep{baraffe98}.

%
%
%
%
\subsection{New photometric candidates}
\label{Pleiades:new_cand_photONLY}

The GCS provides a larger areal coverage than the CFHT and INT
surveys combined (Fig.\ \ref{fig_Pleiades:coverage}).
As a consequence, there are additional candidates lying on
the sequence defined by high probability members (p$\geq$0.6)
in the ($Z-J$,$Z$) CMD (Fig.\ \ref{fig_Pleiades:ZJZcmd}) for 
which no proper motion is available because of a lack of optical
coverage. In this section we investigate their membership using 
the five $ZYJHK$ photometric bands available from the GCS\@.
Those objects are defined thereafter as new photometric
candidates and are included in the computation of the
binary frequency but not in the derivation of the
luminosity and mass functions.

We have considered a straight line passing below the
sequence defined by high probability members in the
($Z-J$,$Z$) CMD (Fig.\ \ref{fig_Pleiades:ZJZcmd}) and
shifted it downwards by 0.2 mag to take into account
photometric errors and the depth of the cluster.
This line goes from ($Z-J$,$Z$) = (1.0, 16.0) to
(2.55, 21.50) and the selection returned 230 candidates.
Among them are the 46 high probability members (including 
three classified as photometric non-members; 
Table \ref{tab_Pleiades:ZJcand}) as well as 34
low probability objects (p$<$60\%), yielding a total of
230$-$46$-$34 = 150 new photometric candidates fainter than
$Z$ = 16 mag. The first step consisted in cross-correlating them 
with the list of 2MASS sources brighter than $J$ = 15.8 mag
to evaluate their membership. Indeed, we have extracted 100 sources 
in 2MASS, including 53 with PM inconsistent with the Pleiades
(outside a circle of radius 25 mas/yr centered on the cluster
mean motion; Table \ref{tab_Pleiades:ZJcand_PM_NM}), 
leaving 47 PM and 50 photometric candidates 
for further investigation. The radius of 25 mas/yr corresponds
to a detection greater than 3$\sigma$ for the birght sources
and 2$\sigma$ for the faintest ones, assuming uncertainties
on the proper motions derived from the cross-correlation
between the GCS and previous optical surveys 
(Sect.\ \ref{Pleiades:new_cand_INT_CFHT}).
Inspection of the location of these new candidates in several CMDs, 
including ($Y-J$,$Y$) and ($J-K$,$J$), have revealed 24 of them 
as photometric non-members (Table \ref{tab_Pleiades:ZJcand_photNM}).
The rejection is based on the blue colors observed in several
diagrams compared to the sequence of PM and photometric members
drawn in Sect.\ \ref{Pleiades:new_cand_proba}.
Consequently, we have 73 new photometric candidates
(Table \ref{tab_Pleiades:ZJcand}) to be added to the 
high probability PM members.

%
%
%
\begin{figure*}
   \centering
   \includegraphics[width=0.49\linewidth]{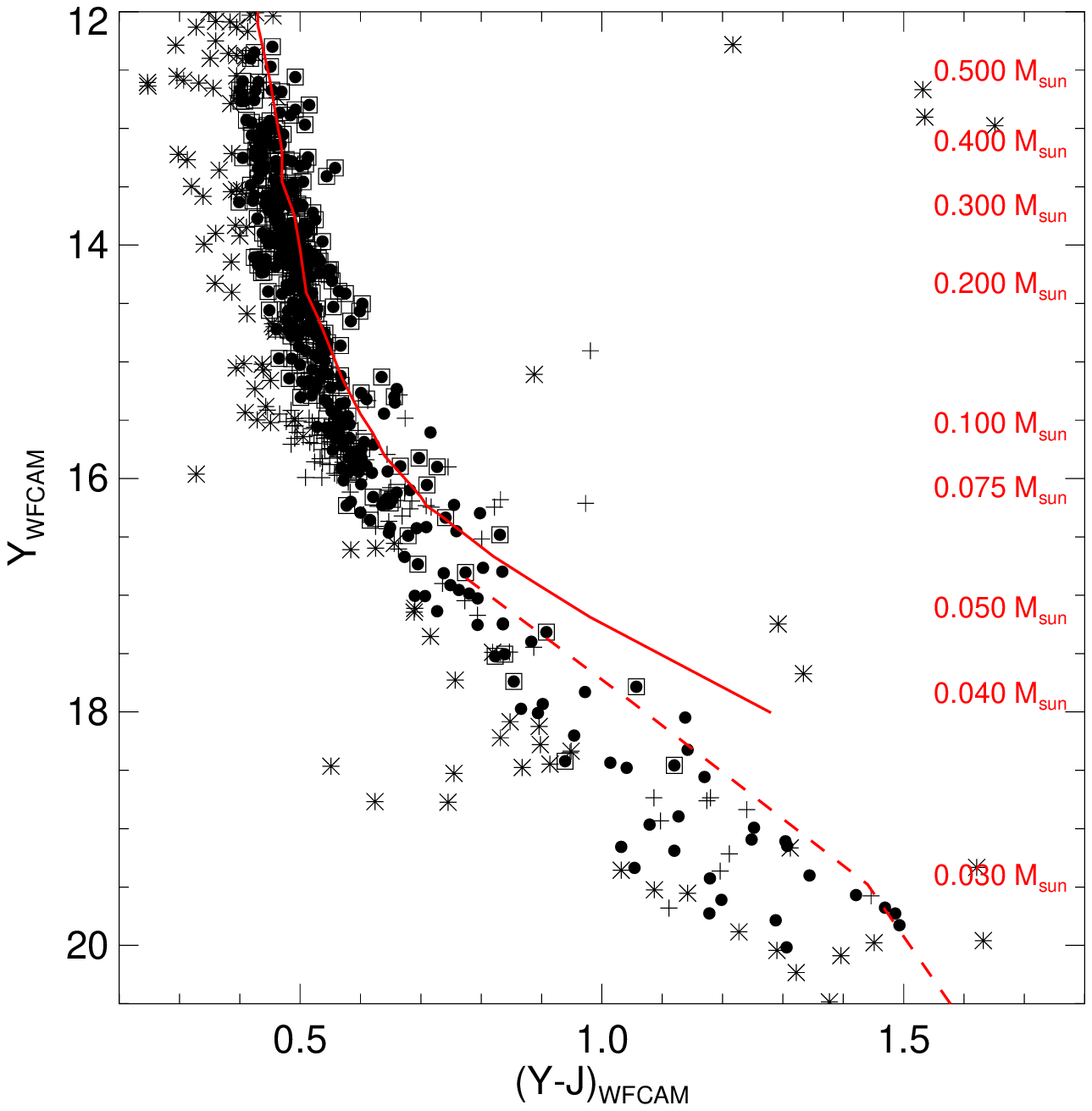}
   \includegraphics[width=0.49\linewidth]{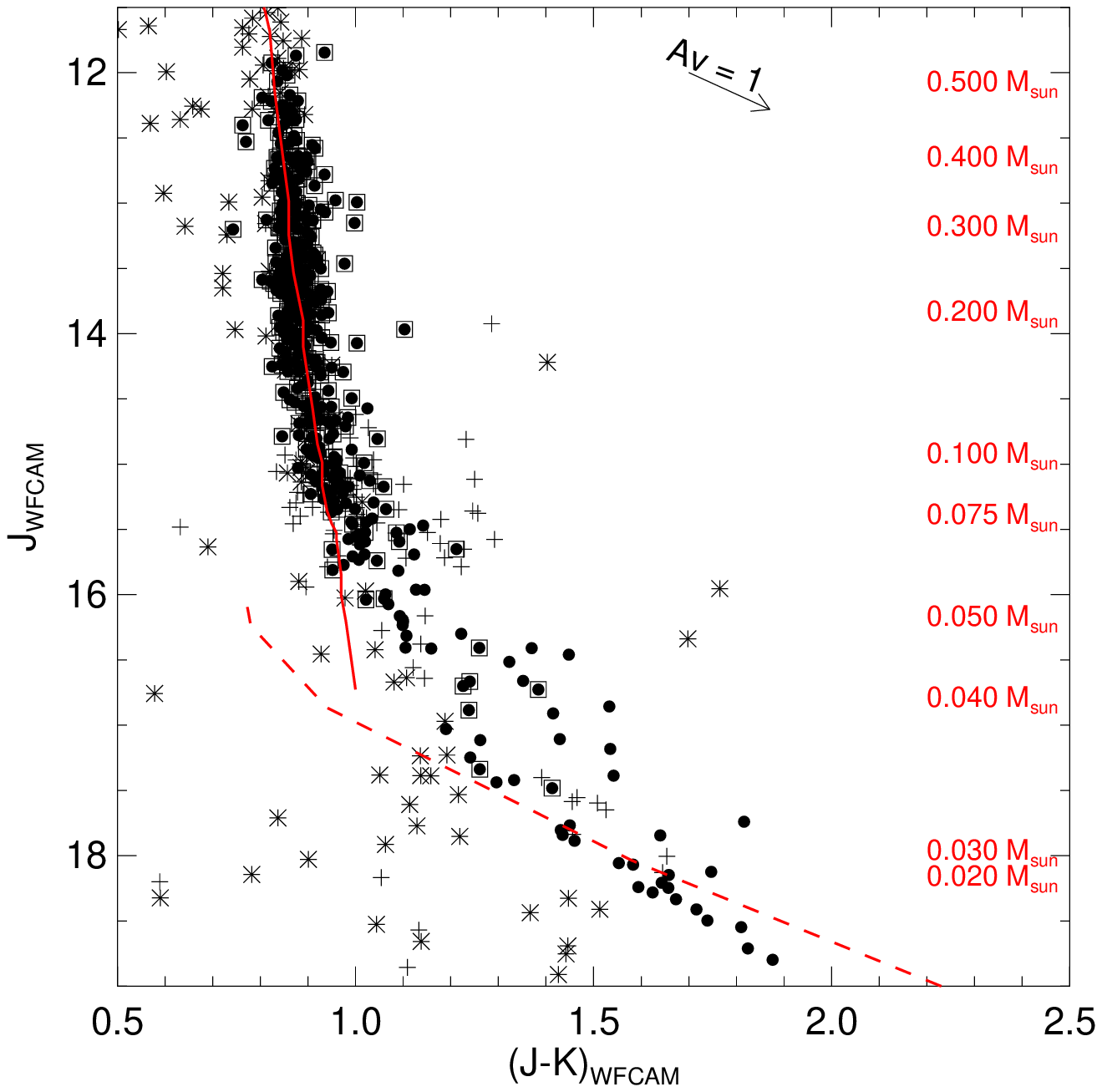}
   \caption{Colour-magnitude diagrams used to extract new low-mass
   Pleiades BD candidates: {\it{left:}} ($Y-J$,$Y$)
   and {\it{right:}} ($J-K$,$J$). All high-probability PM
   members are shown as filled dots with open squares. Other symbols
   are as follows: filled dots are new photometric candidates,
   star symbols are photometric non-members, and crosses are
   PM non-members.
   Overplotted are the 120 Myr NextGen \citep[solid line;][]{baraffe98}
   and DUSTY \citep[dashed line;][]{chabrier00c} isochrones.
   The mass scale is shown on the right hand side of the diagrams
   and spans 0.5--0.03 M$_{\odot}$, according to the
   120 Myr isochrone models.
}
   \label{fig_Pleiades:YJK_cmds}
\end{figure*}

%
%
\subsection{New faint $YJHK$ candidates}
\label{Pleiades:new_cand_faintYJ}

In addition, we have searched for fainter BD candidates using 
the ($Y-J$,$Y$) CMD (left panel in Fig.\ \ref{fig_Pleiades:YJK_cmds};
Table \ref{tab_Pleiades:YJcand}) by imposing $Z$ non detection. 
We have applied the following criteria: 
\begin{itemize}
\item No $Z$ detection
\item $Y$ = 18.5--21.0 mag
\item Candidates should lie above the line defined by 
($Y-J$,$Y$) = (0.75, 18.50) and (1.5, 21.0)
\end{itemize}
This selection has returned 35 additional candidates 
(Table \ref{tab_Pleiades:YJcand}). Inspection of the $YJHK$-band 
images has revealed 9 of them as false or dubious detections 
(cross-talk, discrepant full-width-half-maximum, lack of detection, 
etc\ldots{}), leaving 26 sources for further investigation.
Among them, nine have a proper motion from the cross-correlation
between the (INT$+$CFHT) surveys and the GCS and five are located
within a circle of 25 mas/yr radius centered on the Pleiades mean
motion. The probabilistic approach was not feasible due to the small 
number of $Z$ non-detection with PM\@. All five photometric and
PM members have $J-K$ colours consistent with cluster membership.
By the same token, we note that two out of four PM non-members
were too blue in $J-K$ whereas the other two sit on the cluster
locus and would have been considered as photometric candidates 
if no PM was available. 

Among the remaining 26$-$9 = 17 sources with no PM, seven fit the
cluster sequence in the ($J-K$,$J$) CMD while the remaining
ten have $J-K$ colours bluer than 1.5 mag.
The analysis of the seven sources with PM which also fit the 
cluster sequence in the ($J-K$,$J$) CMD showed that two of them
were classified as PM non-members, implying that would expect at
least two contaminants among the new sample of seven photometric
candidates. Unfortunately, we can only use a statistical 
approach here, hence their membership should
be treated with caution until PM is available for them after
the second observations planned by the GCS\@.

%
%
\subsection{New faint $JHK$ candidates}
\label{Pleiades:new_cand_faintJK}

Finally, we have attempted to select $JHK$-only sources from
the ($J-K$,$J$) CMD to look for even lower-mass BDs
(left panel in Fig.\ \ref{fig_Pleiades:YJK_cmds};
Table \ref{tab_Pleiades:JKcand}). We have imposed
an upper limit of $J$ = 18.9 mag, corresponding to the 100\%
completeness of the GCS\@. Our aim was to avoid any bias towards
red sources as we might reach the L/T transiton predicted at
$J \simeq$ 18.2 mag by the DUSTY models \citep{chabrier00c}.
Hence, we have cross-correlated all GCS sources in the 
$J$ = 17.5--18.9 magnitude range with the INT and CFHT catalogues.
We have retrieved 16 photometric sources with PM
but only one has a PM 
consistent with the Pleiades. Scrutiny of the finding charts
revealed that only three objects were real detections; the
remaining 13 being classified as dubious in
Table \ref{tab_Pleiades:JKcand}.

If a true member and assuming it is a single star, the
faintest photometric and PM candidate in the Pleiades extracted 
from the GCS has $J \simeq$ 18.8 mag and a $J-K$ colour
of 1.88, corresponding to a mass of 28 M$_{\rm Jup}$ according
to the DUSTY models \citep{chabrier00c}. We estimate its
spectral type to be late-L assuming typical near-infrared colours 
for field L dwarfs \citep{vrba04}: the reddest mean $J-K$ colour
is reported for a L8 field dwarf\@. Hence, the GCS 
is just short of the transition region where dust settles in
the atmosphere of BDs. 

Finally, we should mention that the DUSTY models \citep{chabrier00c}
lie on the high side of the Pleiades sequence in the ($Z-J$,$Z$)
(Fig.\ \ref{fig_Pleiades:ZJZcmd}) and ($Y-J$,$Y$) (left panel of 
Fig.\ \ref{fig_Pleiades:YJK_cmds}) CMDs. This fact is not surprising
as the authors stated themselves in their Section 2 that the
DUSTY models represent ``extreme situations'' i.e.\ the amount
of dust could be overestimated if the dust if not in equilibrium 
with the gas phase in brown dwarf's photospheres. 
We find that if we assume that the DUSTY models match the binary
sequence (see Section \ref{Pleiades:binary} for the selection of 
photometric binaries), the errors on the colours and magnitudes could 
be as high as 0.2 and 0.75 mag, respectively. These errors translate
into uncertainties of 200--400 K on the effective temperatures
and 0.01--0.02 M$_{\odot}$ in masses, depending on the mass
of the object.
Furthermore, the DUSTY models cross the cluster sequence in
the ($J-K$,$J$) CMD (right panel of Fig.\ \ref{fig_Pleiades:YJK_cmds}),
suggesting that masses might be underestimated for BDs more massive
than $\sim$0.04 M$_{\odot}$ and underestimated for lower masses.
However, no models can currently reproduce M to L spectral type 
transition satisfactorily and several group are working on this
issue (France Allard, personal communication).

%
%
\section{Cross-correlation with previous surveys}
\label{Pleiades:status_old_cand}

The selection of cluster candidates described above
yielded a large number of previously published Pleiades members
reported in the literature (references therein).
In particular, all bright members down to $Z$ = 16 mag
(corresponding to masses of $\sim$0.1 M$_{\odot}$) reported
in previous survey such as \citet{hambly93}, \citet{adams01a},
and \citet{deacon04}, were recovered by our study.
In Table \ref{tab_Pleiades:early_summary} we summarise the
numbers of proper motion members, photometric candidates,
proper motion and photometric low-mass stars and BD non-members 
published by earlier studies
\citep{festin98,bouvier98,zapatero99b,hambly99,pinfield00,deacon04,bihain06}.
Note that the list published by \citet{deacon04} contains only
sources with membership probabilities higher than 0.6 and
are recovered in the GCS as such.

We provide an electronic table which summarises the information
on previous Pleiades candidates published by earlier studies
and recovered in the GCS area
(Table \ref{tab_Pleiades:electronic_table}). 
This table lists the coordinates 
(J2000), the GCS photometry in 5 passbands ($ZYJHK$), PMs, their 
membership probabilities, as well as their associated names
(including those attributed by previous surveys to the best
of our knowledge).

%
%
\begin{table}
  \caption{Summary of the numbers of photometric and PM members,
photometric candidates only, PM non-members, and photometric members
published by earlier studies and recovered in the GCS area.
The quoted percentages include PM and photometric members
compared to the total number of sources in the GCS area.
References are: \citet{festin98}, \citet{bouvier98}, 
\citet[][ZO99]{zapatero99b}, \citet{hambly99}, \citet{pinfield00}, 
\citet[][DH04]{deacon04}, and \citet{bihain06}.
}
  \label{tab_Pleiades:early_summary}
  \begin{tabular}{l c c c c c r}
  \hline
Survey  & Memb  & PM\_NM & photNM & cand & All &   \%   \cr
  \hline
Festin98    &  14   &  2  &  2    &   4  & 22  &  72.7  \cr
Bouvier98   &  15   & --- &  7    &   1  & 24  &  62.5  \cr
ZO99        &   8   &  4  &  3    &  12  & 24  &  62.5  \cr
Hambly99    &   4   & --- &  1    &  --- &  5  &  80.0  \cr
Pinfield00  & 160   & --- & 45    &  --- & 205 &  78.0  \cr
Moraux03    &  11   & --- &  6    &   6  & 20  &  55.0  \cr
DH04        & 275   & --- &  0    &  --- & 275 & 100.0  \cr
Bihain06    &  20   &  1  &  2    &   1  & 24  &  87.5  \cr
 \hline
\end{tabular}
\end{table}

%
%
%
%
\begin{table*}
  \caption{Sample of objects from the electronic table: 
we list the equatorial coordinates (J2000), GCS $ZYJHK$
photometry, proper motions (mas/yr), names from the previous studies in 
the cluster, and membership (Memb$\equiv$photometric and PM member;
cand$\equiv$photometric candidate, PM\_NM$\equiv$ PM non-member,
photNM$\equiv$photometric non-member).
}
  \label{tab_Pleiades:electronic_table}
  \begin{tabular}{c c c c c c c c c c c}
  \hline
R.A.\ & Dec.\  &  $Z$  &  $Y$  &  $J$  &  $H$  & $K$ & $\mu_{\alpha}cos\delta$ & $\mu_{\delta}$ & Name & Memb? \cr
 \hline
03 54 01.43 & 23 49 57.6 & 99.999   & 99.999   & 18.704   &   17.696 &   16.984 &    ---    &    ---    & BRB29             & cand   \cr
03 54 02.55 & 24 40 25.9 & 20.435   & 19.257   & 18.369   &   17.945 &   17.142 &   10.19   &  $-$20.53 & CFHTPLIZ34        & photNM \cr
 \ldots{}   & \ldots{}   & \ldots{} & \ldots{} & \ldots{} & \ldots{} & \ldots{} & \ldots{}  & \ldots{}  & \ldots{}          & \ldots{} \cr
03 54 14.06 & 23 17 52.0 & 19.946   & 18.760   & 17.586   &   16.789 &   16.131 &   26.13   &  $-$56.27 & CFHTPLIZ28,BRB18  & Memb   \cr
03 54 38.37 & 23 38 01.1 & 20.684   & 19.314   & 18.480   &   17.908 &   17.366 &   $-$5.85 &    4.61   & CFHTPLIZ36        & PM\_NM  \cr
 \hline
\end{tabular}
\end{table*}
%

%
%
%
\begin{figure*}
   \centering
   \includegraphics[width=0.49\linewidth]{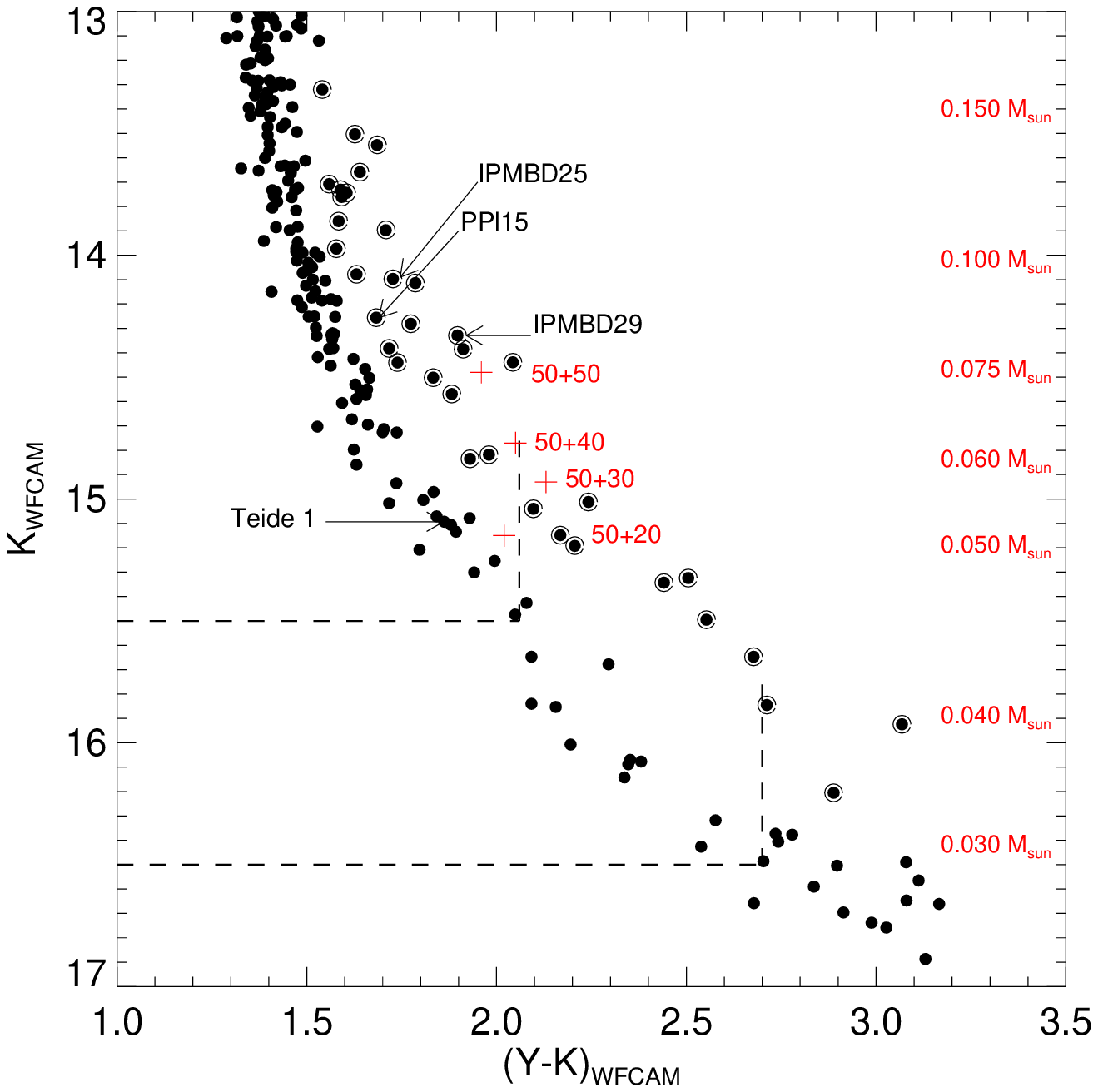}
   \includegraphics[width=0.49\linewidth]{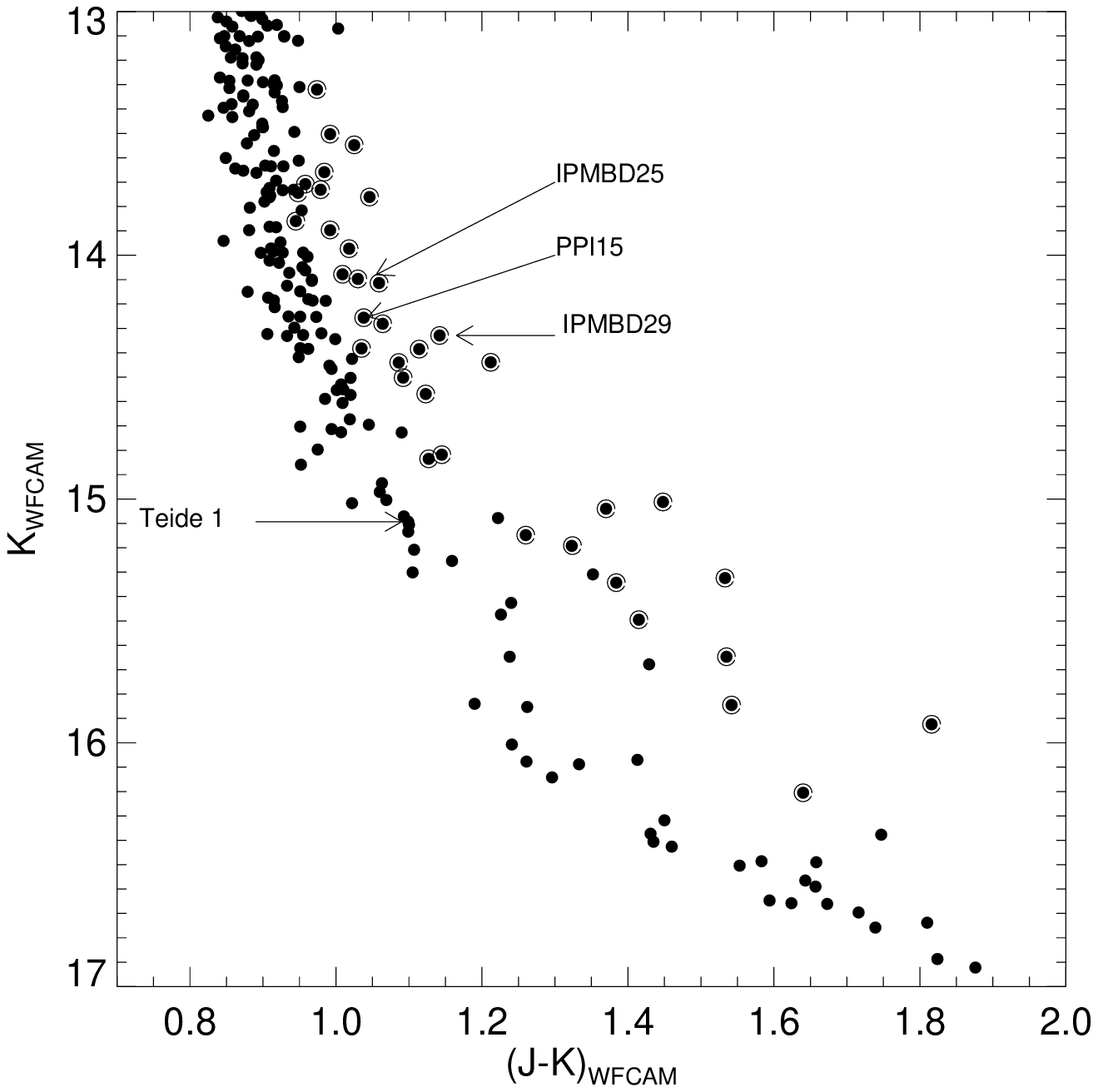}
   \caption{
($Y-K$,$K$) and ($J-K$,$K$) CMDs for all
substellar Pleiades candidates extracted from the UKIDSS GCS\@.
We have originally used the ($Y-K$,$K$) diagram to pick out
substellar photometric multiple system candidates highlighted with 
an open circle around filled dots. We have confirmed all of them 
from their location in the ($J-K$,$K$) diagram. We have found a 
total of 23 photometric multiple systems out of 63 candidates in 
the substellar regime ($K \geq$14.5 mag or masses below 
0.075 M$_{\odot}$), yielding a photometric binary fraction 
of 33-40\%.
We have highlighted a number of known binaries in the
Pleiades, including PPl 15, a spectroscopic binary brown
dwarf with a mass ratio of 0.85 \citep{basri99b},
IPMBD 25, and IPMBD 29 \citep{bouy06a}
as well as Teide 1 which sits in the single star sequence
and for which no companion as been reported to date.
}
   \label{fig_Pleiades:CMD_binary}
\end{figure*}
%

%
%
\section{Discussion on the binary frequency}
\label{Pleiades:binary}

In this section, we investigate the binary frequency of
BDs in the Pleiades using the photometry and colours
from the GCS\@. The multiplicity in the substellar domain
constitutes one way to constrain the formation mechanisms
of BDs.

%
%
\begin{table}
\begin{center}
  \caption{Number of photometric multiple system candidates as
a function of magnitudes extracted from the ($Y-K$,$K$) CMD
and confirmed in the ($J-K$,$K$) diagram
(Table \ref{tab_Pleiades:list_binary}).
Proper motion members and new photometric candidates are included
in the computation of the binary fraction (BF).
Columns list the $K$ magnitude range, the mass range derived for
single stars assuming an age of 120 Myr, the total
number of single stars$+$systems, the number of single stars, the
number of multiple systems (N$_{\rm bin}$), and the binary fraction.
Those calculations suggest a substellar binary frequency
of 28--44\% over the 0.075--0.040 M$_{\odot}$ mass range.
}
  \label{tab_Pleiades:percent_binary}
  \begin{tabular}{c c c c c c}
  \hline
 $K$ range & Mass range    &  All & Single & N$_{\rm bin}$  & BF \cr
  \hline
 \hline
 \hline
14.5--15.5 &  0.075--0.045 &  42  &   27  &  15  &  35.7\% \cr
15.0--16.0 &  0.055--0.040 &  25  &   15  &  10  &  40.0\% \cr
15.5--16.5 &  0.045--0.030 &  21  &   13  &   8  &  38.1\% \cr
14.5--16.5 &  0.075--0.030 &  63  &   40  &  23  &  36.5\% \cr
 \hline
\end{tabular}
\end{center}
\end{table}
The single-star sequence and its associated binary sequence 
lying 0.75 mag above are clearly
visible in the ($Y-K$,$K$) CMD (Fig.\ \ref{fig_Pleiades:CMD_binary}).
We have selected substellar multiple system candidates
from this diagram (circled dots in Fig.\ \ref{fig_Pleiades:CMD_binary})
and listed them in Table \ref{tab_Pleiades:list_binary}.
The multiplicity of those candidates was confirmed in their
position in the ($J-K$,$K$) diagram (right panel in 
Fig.\ \ref{fig_Pleiades:CMD_binary})\footnote{We note that
a few candidates lie in the binary sequence in the ($J-K$;$K$)
CMD but were not selected in the ($Y-K$,$K$) CMD. These
sources could be either lower mass ratio systems or reddened
cluster members}.
The selection was made as follows: for a given magnitude,
say $K$ = 15.5--16.5 mag, we have drawn two horizontal lines
(dashed lines in Fig.\ \ref{fig_Pleiades:CMD_binary})
intercepting the single object sequence which we have interpolated
using high probability PM members available from the
optical/infrared cross-correlation 
(Section \ref{Pleiades:new_cand_PM}). From the intercept 
points we have drawn two vertical lines with a length of 0.75 mag.
Then we have divided the box formed by both sequences and the vertical 
lines into two boxes: single stars lie in the bottom part whereas
binary candidates in the top one. The dividing line between the lower
and upper box was chosen to go through the gap present between the
single-star and binary sequences, corresponding to a mass ratio
of $\sim$0.4\@. The binary fraction was then defined as the
number of binaries divided by the total number of
objects (single stars$+$binaries).

In total, we have counted 23 substellar multiple system candidates 
in the $K$ = 14.5--16.5 mag range (corresponding to masses between 
0.075 and 0.030 M$_{\odot}$) and 40 single objects, yielding
a binary frequency of 23/(23$+$40) = 36.5$\pm$8.0\% assuming 
Poisson errors (Table \ref{tab_Pleiades:percent_binary}). 
These candidates include high probability
members and new photometric candidates.
The binary frequency range across the substellar regime probed
by our survey is more likely to lie between 35 and 40\%
according to the dispersion as a function of mass
(Table \ref{tab_Pleiades:percent_binary}).

Our estimate of the substellar binary frequency
is likely to be a lower limit for two reasons.
Firstly, high-mass ratio binaries i.e. the ones with
very low-mass companions will have hardly moved away from the
single star sequence. For example, the crosses plotted
in the left panel of Fig.\ \ref{fig_Pleiades:CMD_binary} 
represent a 20, 30, 40, and 50 M$_{\rm Jup}$ BD added to a 
50 M$_{\rm Jup}$ BD, respectively.
The computation of the sequence of
binaries with successively more massive companions
was done using the Lyon group models for $K$ magnitudes
and empirical data for $Y$ magnitudes. Adding intensities for
various mass combinations and reconverting to magnitudes
gave the crosses in Fig.\ \ref{fig_Pleiades:CMD_binary}.
It can be seen that low-mass ratio binary systems
will hardly stand out from the single star sequence,
hence difficult to pick out in a photometric-based search.
As a consequence of the choice of our dividing line, we
should be sensitive to mass ratios greater than 0.4
as are high-resolution imaging surveys.
Secondly, the single star sequence is more likely to be
affected by field dwarfs and reddened background stars
than the binary sequence.
Additionally, we need to take into account the depth
of the cluster. The Pleiades has a tidal radius of
13 pc \citep{pinfield00} and a distance of 130 pc,
implying that the distance of any member can
vary by $\pm$10\% corresponding to a variation
of $\pm$0.2 magnitude.

On average, we have derived a BD binary fraction of 28-44\% 
in the Pleiades cluster, two to three times larger than estimates
from high-resolution imaging survey 
\citep[13.3$^{+13}_{-4}$\%;][]{martin00a,martin03,bouy06a}. 
The upper limit from the {\it{HST}} surveys in the Pleiades
is lower than our estimates. We probe a wider 
mass range (75--30 M$_{\rm Jup}$ vs 65--55 M$_{\rm Jup}$)
and a comparable mass ratio range (q$\geq$0.5) than those
high-resolution surveys.
Furthermore, our estimate is lower than the photometric
estimate of 50$\pm$10\% by \citet{pinfield00} although
consistent within the uncertainties.
Recently, \citet{basri06} inferred an upper limit of
26$\pm$10\% on the binary fraction of low-mass stars and BDs,
divided up into 11\% of spectroscopic binaries (0--6 au) and
15\% of wider binaries (3--15 au) despite some overlap
between both subsample. Our results are on the high side of
this latter estimate but compatible considering the uncertainties
on both measurements. Finally, our results are in excellent 
agreement with the frequency of 32--45\% from Monte-Carlo 
simulations \citep{maxted05}.
Our results are not reproduced by current theoretical models
predicting a low fraction of substellar binaries 
\citep{bate02,delgado_donate03} and represent a challenge for 
current theory of brown dwarf formation.
Moreover we find a gap between the single and binary sequences,
suggesting that most BDs tend to harbour equal-mass ratios
seven sources have q=0.6--0.8; 15 have q=0.8--1.0;
two have q$<$0.6, and twelve are undetermined), similar to the
properties of field BDs (mass ratios are listed in the last
column of Table \ref{tab_Pleiades:percent_binary}).
Furthermore, we have recovered two known binaries resolved by 
{\it{HST}} (IPMBD 25 and 29) and one object observed by {\it{HST}} 
but not resolved (CFHT-Pl-IZ 4) as well as PPl15, suggesting
that high-resolution surveys could be missing half of the
binaries, hypothesis consistent with the conclusions drawn
by \citet{basri06}. Finally, the binary fraction could be 
dependent on the environment with a possible higher frequency
in clusters than the field 
\citep{pinfield00,maxted05,kraus06,bouy06b,burgasser07a},
trend reproduced by a model proposed by \citet{goodwin07}.
However, there is currently no direct evidence for such a
dependence mainly due to the lack of statistics and the
uncertainties on membership in young clusters 
\citep{kraus06,bouy06b}.

%
%
\section{The Initial mass function}
\label{Pleiades:IMF}

In this section we discuss the cluster luminosity and
mass function derived from the sample of candidates with 
PM from the ``bright'' (2MASS) and ``faint'' (INT$+$CFHT) samples. 
We do not consider here the new photometric candidates
(Section \ref{Pleiades:new_cand_photONLY}) because
of a lack of PMs and did not attempt to correct the mass
function for binaries.

%
%
%
\begin{figure*}
   \centering
   \includegraphics[width=0.49\linewidth]{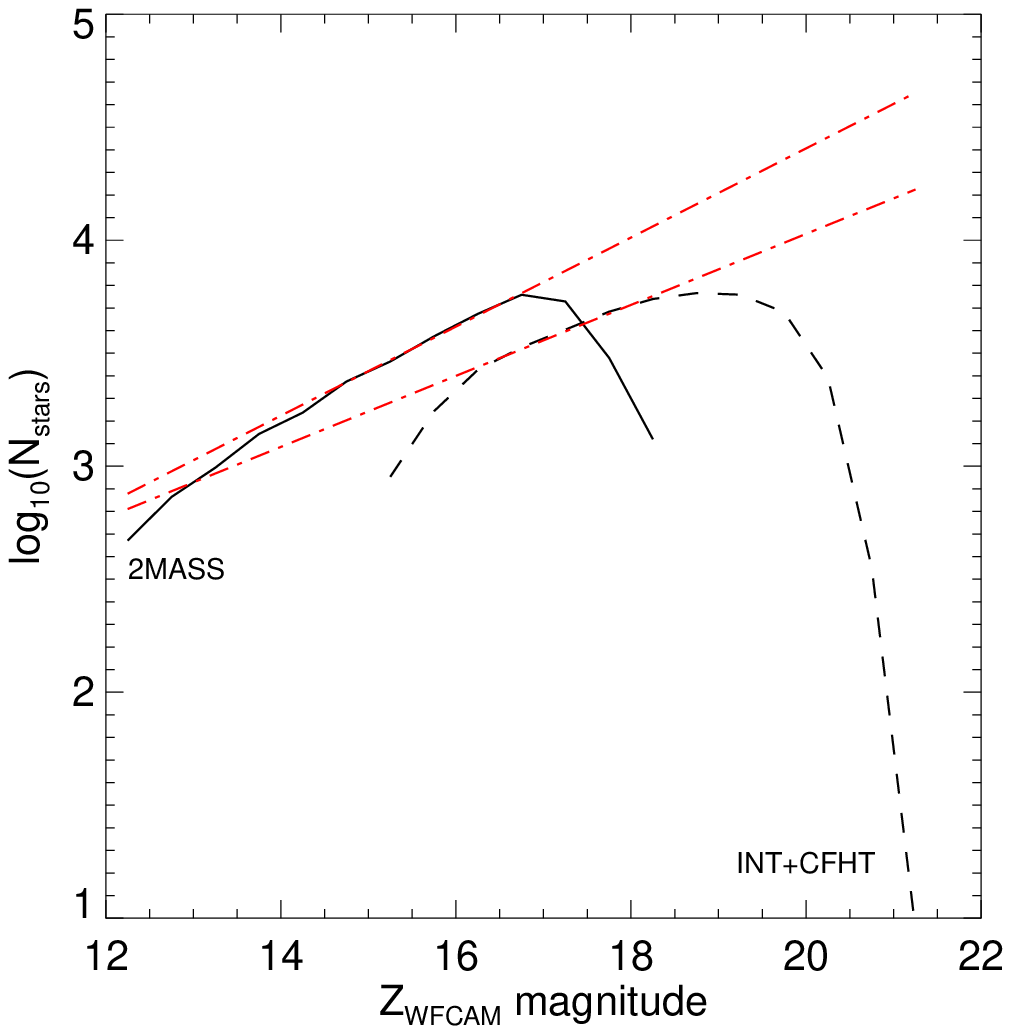}
   \includegraphics[width=0.49\linewidth]{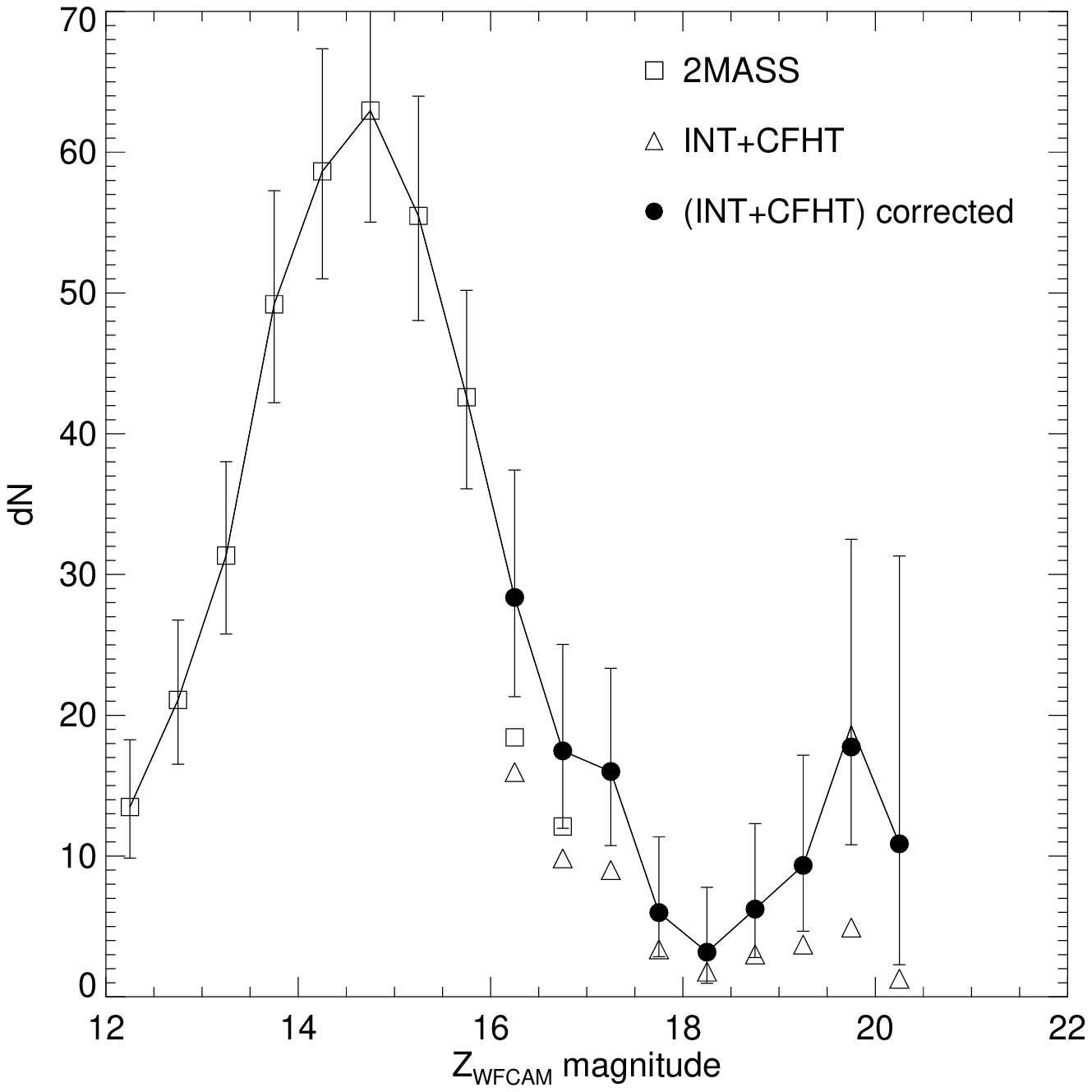}
   \caption{
{\it{Left:}} The number of stars as a function of magnitude for
the ``bright'' (2MASS; solid line) and ``faint'' (INT$+$CFHT; dashed
line) samples. The red dot-dashed lines represent the linear fit to
both sets of data and are used to estimate their incompleteness.
Both samples are complete in the $Z$ = 16.25--16.75 magnitude range,
implying a scaling factor of 1.777\@.
{\it{Right:}} The Pleiades luminosity function, defined as the sum
of all membership probabilities per 0.5 magnitude bins, for all
candidates in the $Z$ = 12.0--21.5 magnitude range.
The ``bright'' sample is shown as open squares whereas the ``faint''
sample is depicted as open triangles. The filled circles
represent the ``faint'' sample corrected for incompleteness
and scaling. The last two points are not plotted because outside
the plot scale. Error bars are Gehrels error bars.
}
   \label{fig_Pleiades:LF_plot}
\end{figure*}
%

%
%
\begin{table*}
  \caption{Luminosity and mass function for the Pleiades
  open cluster using the NextGen and DUSTY 120 Myr theoretical
  isochrones and a distance of 130 pc.
  The incompleteness factor (InFactor) applied to the faint
  end of the luminosity function are quoted as well as the
  scaling factor (ScFactor) of 1.777 applied to the ``faint''
  sample to take into account the limited optical coverage
  from the INT and CFHT surveys and match the ``bright'' sample.
  Gehrels errors, scaled by the incompleteness and scaling
  factors, are listed for the luminosity and mass functions
  (errH and errL stand for upper and lower error bars, respectively).
}
  \label{tab_Pleiades:LF_MF}
  \begin{tabular}{c c c c c c c c c c c}
  \hline
Mag range &   Mass range  & Mid-mass & dN   & errH & errL & InFactor & ScFactor & dN/dM & errH & errL \cr
  \hline
12.0-12.5 & 0.6470-0.5630 & 0.60500 &  13.5 &  4.8 &  3.6 &    1.00 & 1.000 &    160.6 &     56.8 &     43.3 \cr
12.5-13.0 & 0.5630-0.4940 & 0.52850 &  21.1 &  5.7 &  4.6 &    1.00 & 1.000 &    305.7 &     82.2 &     66.2 \cr
13.0-13.5 & 0.4940-0.4170 & 0.45550 &  31.4 &  6.7 &  5.6 &    1.00 & 1.000 &    407.2 &     86.6 &     72.4 \cr
13.5-14.0 & 0.4170-0.3330 & 0.37500 &  49.2 &  8.1 &  7.0 &    1.00 & 1.000 &    585.7 &     96.0 &     83.3 \cr
14.0-14.5 & 0.3330-0.2520 & 0.29250 &  58.6 &  8.7 &  7.6 &    1.00 & 1.000 &    724.0 &    107.5 &     94.3 \cr
14.5-15.0 & 0.2520-0.1920 & 0.22200 &  63.0 &  9.0 &  7.9 &    1.00 & 1.000 &   1049.2 &    149.7 &    132.0 \cr
15.0-15.5 & 0.1920-0.1460 & 0.16900 &  55.5 &  8.5 &  7.4 &    1.00 & 1.000 &   1206.0 &    184.7 &    161.6 \cr
15.5-16.0 & 0.1460-0.1160 & 0.13100 &  42.6 &  7.6 &  6.5 &    1.00 & 1.000 &   1420.0 &    252.8 &    216.9 \cr
16.0-16.5 & 0.1160-0.0920 & 0.10400 &  28.4 &  6.4 &  5.3 &    1.00 & 1.777 &   1182.2 &    266.5 &    221.0 \cr
16.5-17.0 & 0.0920-0.0750 & 0.08350 &  17.5 &  5.3 &  4.2 &    1.00 & 1.777 &   1027.8 &    309.9 &    244.1 \cr
17.0-17.5 & 0.0750-0.0645 & 0.06980 &  16.0 &  5.1 &  4.0 &    1.00 & 1.777 &   1524.5 &    485.1 &    378.1 \cr
17.5-18.0 & 0.0645-0.0558 & 0.06020 &   6.0 &  3.6 &  2.4 &    1.00 & 1.777 &    687.7 &    413.2 &    275.2 \cr
18.0-18.5 & 0.0558-0.0495 & 0.05270 &   3.2 &  3.0 &  1.7 &    1.00 & 1.777 &    504.0 &    473.2 &    271.5 \cr
18.5-19.0 & 0.0495-0.0416 & 0.04560 &   6.2 &  3.6 &  2.4 &    1.16 & 1.777 &    789.0 &    461.1 &    309.6 \cr
19.0-19.5 & 0.0416-0.0377 & 0.03960 &   9.3 &  4.2 &  3.0 &    1.43 & 1.777 &   2394.1 &   1070.8 &    772.9 \cr
19.5-20.0 & 0.0377-0.0349 & 0.03630 &  19.3 &  5.5 &  4.4 &    2.03 & 1.777 &   6881.4 &   1955.0 &   1557.5 \cr
20.0-20.5 & 0.0349-0.0321 & 0.03350 &  10.9 &  4.4 &  3.3 &    4.74 & 1.777 &   3882.9 &   1574.7 &   1164.0 \cr
20.5-21.0 & 0.0321-0.0298 & 0.03095 &  72.9 &  9.6 &  8.5 &   39.29 & 1.777 &  31695.7 &   4166.1 &   3705.9 \cr
21.0-21.5 & 0.0298-0.0289 & 0.02935 & 268.1 & 17.4 & 16.4 & 1862.86 & 1.777 & 297926.7 &  19330.8 &  18185.7 \cr
 \hline
\end{tabular}
\end{table*}
\subsection{The cluster luminosity function}
\label{Pleiades:IMF_LF}

In this section, we consider all photometric candidates with
membership probabilities i.e.\ 1061 sources with $Z$ magnitudes 
ranging from 12 to 21.5 mag (Section \ref{Pleiades:new_cand_photONLY}). 
Assuming that the lithium depletion boundary is at 
M$\sim$0.075 M$_{\odot}$ \citep[M$_{Z}$ = 11.44][]{stauffer98,barrado04b}
and a distance of 130 pc, our sample contains 967 stars and 94 BDs. 
Those numbers objects should not be seen as exact numbers as they 
are subject to a 0.15 mag uncertainty in the position of the lithium
depletion boundary but imply that our sample contains
10$^{+1.6}_{-0.5}$\% of substellar objects (the number of BDs 
varies from 82 to 98).
The luminosity function is usually defined as the
number of stars as a function of magnitude but here we have 
derived it in a probabilistic manner: we summed the
membership probability of all 1061 sources divided up into 
0.5 magnitude bin instead of simply counting the number of objects. 
We have two independent samples: on the one hand, the ``bright''
sample considered down to $Z$ = 16 mag and the `faint'' sample 
used for fainter candidates. Both samples are complete over
the $Z$ = 16.25--16.75 magnitude range as demonstrated in
the left-hand side panel of Fig.\ \ref{fig_Pleiades:LF_plot}.

The ``faint'' sample was not drawn from the full GCS sample 
(hence coverage)
because of the limited areal optical coverage and, thus, need
to be scaled to the ``bright'' sample. We have investigated
two approaches to infer a scaling factor over the
$Z$ = 16.25--16.75 magnitude range where both samples were
complete. The total number of sources with proper motions 
in this magnitude range is 5988 and 2976 for the bright and 
faint samples, respectively. Hence, the ratio of the number 
of sources gives a scaling factor of 5288/2976 = 1.777\@. 
The second consisted in summing the probabilities for
the ``bright'' and ``faint'' samples in the same magnitude
range as above. The inferred scaling factor is 20.217/12.114 = 1.669\@.
Hence, we have favoured the first method (which is likely less
affected by uncertainties) and scaled the luminosity function of the 
``faint'' sample by a factor of 1.777 to match the ``bright'' 
sample. The error on the scaling factor is less than 0.1 since 
both methods give similar results.

The sum of membership probabilities after scaling and correction
for incompleteness are given in Table \ref{tab_Pleiades:LF_MF} and 
plotted in the right-hand side panel of 
Fig.\ \ref{fig_Pleiades:LF_plot}. Error bars 
are Gehrels errors \citep{gehrels86} rather than Poissonian error 
bars because the former represent a better approximation to the 
true error for small numbers. The upper limit is defined as
1+($\sqrt(dN+0.75)$) and the lower limit as $\sqrt(dN-0.25)$
assuming a confidence level of one sigma.
Incompleteness and scaling factors were applied to the errors
when necessary.
The luminosity function peaks at $Z \sim$ 14.5--15.0 mag and 
decreases down to $Z \sim$ 18.0--18.5 mag where it bounces 
back down to the survey limit (right-hand side panel of
Fig.\ \ref{fig_Pleiades:LF_plot}). 
The last two bins are subject to large incompleteness factors
and should be treated with caution. Similarly, the brightest
bin is likely to be affected by incompleteness.

\subsection{The cluster mass function}
\label{Pleiades:IMF_MF}

In this section we adopt the following definition for the
mass function as originally proposed by \citet{salpeter55}:
$\xi$($\log_{10}m$) = d$n$/d$\log_{10}$($m$) $\propto$ m$^{-\alpha}$.
We have converted the luminosity into a mass function using 
the NextGen models \citep{baraffe98} for stars and BDs
more massive than 50 M$_{\rm Jup}$ (T$_{\rm eff}$) and the DUSTY 
models \citep{chabrier00c} below. The $Z$ = 12--21.5 magnitude 
range translates into masses between 0.65 and 0.03 M$_{\odot}$, 
assuming a distance of 130 pc and an age of 120 Myr.
This mass range is in agreement with the estimated masses
given in Table 2 in \citet{schwartz05}, suggesting that we
have detected mid-L dwarfs in the Pleiades.
The latter authors used the DUSTY models \citep{chabrier00c}
and Burrows models \citep{burrows97,burrows00} at 80 and 125 Myr 
to infer a possible mass range for their new BD candidates.
The faintest photometric and PM candidate extracted from
the ($J-K$,$J$) CMD but not included in this mass function
exhibit red infrared colours (Section \ref{Pleiades:new_cand_faintJK}),
implying that we are not probing the L/T transition in the
Pleiades where dust settles in the atmosphere of BDs.

\begin{figure*}
   \centering
   \includegraphics[width=\linewidth]{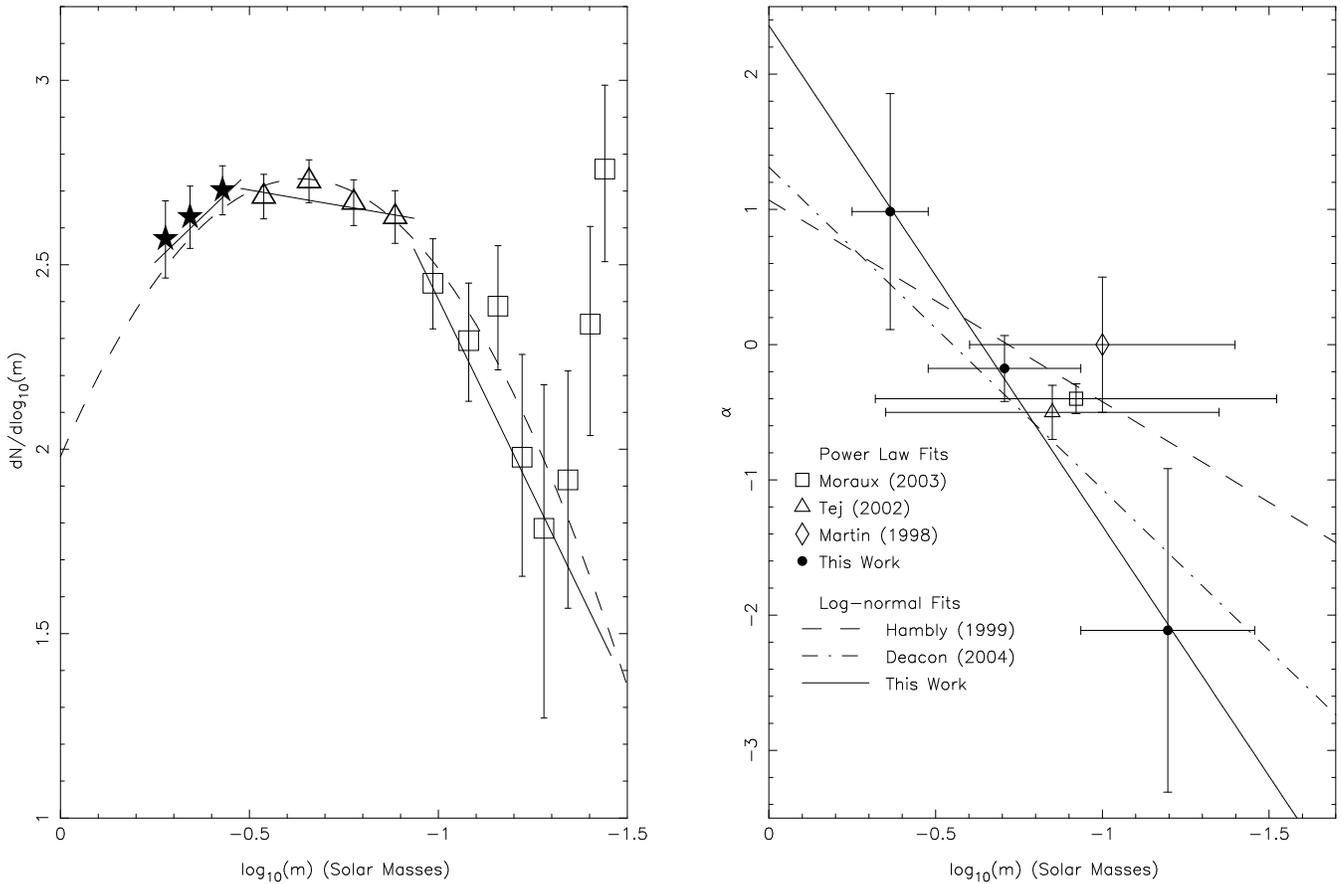}
   \caption{
{\it{Left:}} The Pleiades mass function for all candidates
extracted from the UKIDSS GCS DR1\@. We have assumed a distance of
130 pc and an age of 120 Myr for the Pleiades cluster.
The NextGen and DUSTY models were used to transform magnitudes
into masses. Error bars are Gehrels errors.
Three power law segments (solid lines) were fit to the mass
function with $\alpha$ parameters as follows: 0.98$\pm$0.87
in the 0.563--0.333 M$_{\odot}$ mass range (filled circles),
$-$0.18$\pm$0.24 for 0.333-0.116 M$_{\odot}$ (star
symbols), and $-$2.11$\pm$1.20 over 0.116-0.035 M$_{\odot}$
(open squares). The lognormal function (dashed lines)
has the following parameters $a0$ = 1.99, $a1$ = $-$2.37, and
$a2$ = $-$1.89 (using the definition in Section \ref{Pleiades:IMF_MF}).
{\it{Right:}} The $\alpha$-plot for the Pleiades mass function.
The filled circles represent the three power law segments from
our study and the solid line the lognormal function fit
to the mass function. Other symbols and lines depict previous
studies: \citet[][; open diamond]{martin98a},
\citet[][dashed line]{hambly99},
\citet[][open diamond]{martin98a},
\citet[][open triangle]{tej02},
\citet[][open square]{moraux03}, and
\citet[][dot-dashed line]{deacon04}.
Vertical segments represent the error on the $\alpha$ index
whereas the horizontal bars represent the range in mass
where the $\alpha$ index is valid.
}
   \label{fig_Pleiades:MF_plot}
\end{figure*}

The mass function, derived from the sample of sources with 
membership probabilities, is shown in the left-hand side panel 
of Fig.\ \ref{fig_Pleiades:MF_plot}. This mass function
is the ``unresolved system'' mass function since we did not
attempt to correct for binaries \citep{moraux04}. Similarly
we did not correct for the radial distribution of low-mass
stars and BDs in the cluster due to the inhomogeneous
coverage currently available from the GCS\@. On inspection 
of the mass function (Fig.\ \ref{fig_Pleiades:MF_plot}) it was 
decided that a single power law would not properly represent 
the functional form over the full range of masses. Hence we 
have considered three
mass ranges where we have fitted a power law: 
0.563--0.333 M$_{\odot}$ (filled circles),
0.333--0.116 M$_{\odot}$ (star symbols), and
0.116--0.035 M$_{\odot}$ (open squares).
Each segment in the mass range is best characterised by power 
law indices $\alpha$ of 0.98$\pm$0.87, $-$0.18$\pm$0.24,
$-$2.11$\pm$1.20, respectively. Those results are in
agreement within the uncertainties with previous studies
of the Pleiades mass function in the low-mass star and brown 
dwarf regimes \citep{martin98a,dobbie02a,tej02,moraux03,deacon04}.
\citet{adams96} proposed that the initial mass function
can be approximated by a lognormal function defined by:
\begin{equation}
\log_{10} \xi(\log_{10}m) = a_0 + a_1 \times \log_{10} m + a_2 \times (\log_{10} m)^2
\end{equation}
The best lognormal function fit to the mass function (dashed line
in the left hand side panel of Fig.\ \ref{fig_Pleiades:MF_plot})
is given by the following parameters:
$a_0$ = 1.99, $a_1$ = $-$2.37, and $a_2$ = $-$1.89 over the 
0.56--0.035 M$_{\odot}$ mass range. These parameters can be 
converted into those defined by \citet{chabrier03b} to yield 
$m_c \sim 0.24$ and $\sigma \sim$ 0.34\@. The value of $m_c$ is in good 
agreement with the value of 0.22 found by \citet{chabrier03b} 
for a mass function which includes unresolved multiple systems. 
However the value of $\sigma$ (0.59) does differ from our results.
How significant this difference is is impossible to say as Chabrier
did not publish errors on his parameters.

It is clear that mass function deviates from a single power law 
over the course of our sample, with a peak around $0.24M_{\odot}$. 
In order to compare these results with those from previous studies 
an $\alpha$ plot was produced showing the variation in the value 
of $\alpha$ over the full mass range (right hand side panel
in Fig.\ \ref{fig_Pleiades:MF_plot}). The results from 
several other studies are plotted here too
\citep{martin98a,hambly99,tej02,moraux03,deacon04}. 
It can be seen that the lognormal functions published by 
\citet{hambly99} and \citet{deacon04} are also in reasonable 
agreement with these results. 

%
%
%
%

%
%
\section{Summary}
\label{Pleiades:summary}

We have presented a deep wide-field survey conducted in the Pleiades
open cluster as part of the UKIDSS Galactic Cluster Survey. 
We have employed a photometric selection complemented with a PM 
probabilistic analysis to assess the membership of Pleiades 
member candidates. 
We have recovered bright members from previous photometric and PM 
surveys down to 0.1 M$_{\odot}$. Below this limit, we present a 
catalogue of high probability PM members as well as new photometric 
BD candidates down to 0.03 M$_{\odot}$.
The main outcomes of the GCS study in the central region of the
Pleiades are two-fold:
\begin{itemize}
\item we have derived a BD binary fraction around 33--40\% 
in the 0.075--0.030 M$_{\odot}$ mass range using the sample of
high probability PM members as well as the photometric sample.
This estimate is in agreement with Monte-Carlo simulations of
\citet{maxted05} and on the upper side of the estimate from
the radial velocity survey conducted by \citet{basri06}.
The inferred binary frequency is however on the low side of the 
photometric estimate by \citet{pinfield03} and significantly 
higher than predictions by high-resolution imaging surveys.
The separations and mass ratios seems however consistent with
findings for ultracool field dwarfs and BDs in the Pleiades.
\item we have derived the Pleiades luminosity function using the
sample of photometric candidates with membership probabilities.
We have inferred a mass function in the 0.56--0.03 M$_{\odot}$ 
and fitted a lognormal function peaking at 0.24 M$_{\odot}$ which
in agreement with previous studies in the cluster.
\end{itemize}

%
%
\section*{Acknowledgments}

This work is based in part on data obtained as part of the UKIRT 
Infrared Deep Sky Survey (UKIDSS).
Part of this work was carried out in Leicester where NL and PDD were 
postdoctoral research associates funded by the UK PPARC.
We are grateful to Isabelle Baraffe and France Allard for providing 
us with the NextGen, DUSTY and COND models for the WFCAM filters.
We thank our colleagues at the UK Astronomy Technology Centre,
the Joint Astronomy Centre in Hawaii, the Cambridge Astronomical Survey
and Edinburgh Wide Field Astronomy Units for building and operating
WFCAM and its associated data flow system.
This research has made use of the Simbad database, operated at
the Centre de Donn\'ees Astronomiques de Strasbourg (CDS), and
of NASA's Astrophysics Data System Bibliographic Services (ADS).
This publication has also made use of data products from the
Two Micron All Sky Survey, which is a joint project of the
University of Massachusetts and the Infrared Processing and
Analysis Center/California Institute of Technology, funded by
the National Aeronautics and Space Administration and the National
Science Foundation.

%
%
\bibliographystyle{mn2e}
\bibliography{../../AA/mnemonic,../../AA/biblio_old}

%
%

%
%
\appendix

%
%
%
%
\section{Table of of high probability proper motion members
and new photometric candidates in the Pleiades
from the UKIDSS GCS DR1\@.}

\begin{table*}
  \caption{Near-infrared ($ZYJHK$) photometry for 43 high
probability members and 73 new photometric candidates in
the Pleiades open clusters. All sources are fainter than
$Z$ = 16 mag, corresponding to masses below 0.1 M$_{\odot}$
according to theoretical models.
This table lists the equatorial
coordinates (in J2000), magnitudes from the GCS,
PMs, membership probabilities when available, membership
status, and names from the literature for each source.
(memb stands for high probability (p$\geq$0.6) PM members
whereas cand qualify new photometric candidates.
}
  \label{tab_Pleiades:ZJcand}
  \begin{tabular}{c c c c c c c c c c c l}
  \hline
R.A.\ & Dec.\  &  $Z$  &  $Y$  &  $J$  &  $H$  & $K$ & $\mu_{\alpha}cos\delta$ & $\mu_{\delta}$ & Prob & Memb? & Other name \\
 \hline
03 40 35.50 & 23 13 07.4 & 17.908 & 16.811 & 16.073 & 15.499 & 15.004 &   19.61 &  $-$35.45 &  ---  & cand & UGCS$-$Pl$-$1  \cr
03 40 39.46 & 23 26 34.8 & 16.167 & 15.563 & 15.020 & 14.455 & 14.062 &   16.55 &  $-$41.62 &  ---  & cand & NoName     \cr
03 40 52.79 & 25 24 43.4 & 19.156 & 18.009 & 17.115 & 16.423 & 15.853 &    ---  &    ---  &  ---  & cand & UGCS$-$Pl$-$2  \cr
03 40 55.30 & 25 34 57.4 & 17.381 & 16.464 & 15.817 & 15.187 & 14.727 &   21.07 &  $-$58.68 &  ---  & cand & UGCS$-$Pl$-$3  \cr
03 41 30.36 & 25 17 06.0 & 16.555 & 15.821 & 15.240 & 14.674 & 14.297 &   11.31 &  $-$40.18 & 0.849 & memb & UGCS$-$Pl$-$4  \cr
03 41 33.90 & 23 11 44.9 & 16.134 & 15.557 & 15.029 & 14.501 & 14.150 &   37.63 &  $-$33.57 &  ---  & cand & HHJ12      \cr
03 41 37.74 & 23 04 32.7 & 17.929 & 17.005 & 16.315 & 15.658 & 15.208 &    ---  &    ---  &  ---  & cand & UGCS$-$Pl$-$5  \cr
03 41 40.90 & 25 54 24.1 & 16.829 & 15.900 & 15.173 & 14.562 & 14.114 &   23.78 &  $-$37.48 & 0.915 & memb & CFHT$-$PLIZ4 \cr
03 41 43.72 & 23 07 59.8 & 16.340 & 15.654 & 15.072 & 14.522 & 14.105 &   17.71 &  $-$45.66 &  ---  & cand & UGCS$-$Pl$-$6  \cr
03 41 54.16 & 23 05 04.8 & 17.336 & 16.297 & 15.499 & 14.912 & 14.385 &   10.66 &  $-$24.90 &  ---  & cand & MHOBD3     \cr
03 42 05.74 & 23 07 14.4 & 16.654 & 15.911 & 15.343 & 14.769 & 14.344 &   16.53 &  $-$36.85 &  ---  & cand & UGCS$-$Pl$-$7  \cr
03 42 07.99 & 22 39 33.5 & 18.718 & 17.398 & 16.515 & 15.809 & 15.192 &    ---  &    ---  &  ---  & cand & int$-$pl$-$IZ$-$76 \cr
03 42 47.29 & 23 00 40.4 & 17.009 & 16.194 & 15.554 & 14.965 & 14.553 &   12.99 &  $-$48.09 &  ---  & cand & UGCS$-$Pl$-$8  \cr
03 42 59.92 & 22 42 51.5 & 16.067 & 15.267 & 14.666 & 14.109 & 13.708 &   25.70 &  $-$41.46 & 0.860 & memb & UGCS$-$Pl$-$9  \cr
03 43 11.76 & 25 31 32.0 & 16.020 & 15.352 & 14.778 & 14.233 & 13.897 &    9.82 &  $-$49.41 &  ---  & cand & UGCS$-$Pl$-$10 \cr
03 43 22.55 & 23 00 56.5 & 16.141 & 15.457 & 14.899 & 14.354 & 13.985 &   10.54 &  $-$33.85 &  ---  & cand & UGCS$-$Pl$-$11 \cr
03 43 34.48 & 25 57 30.6 & 16.471 & 15.605 & 14.889 & 14.323 & 13.897 &    9.23 &  $-$40.49 &  ---  & cand & UGCS$-$Pl$-$12 \cr
03 43 56.00 & 25 36 25.3 & 16.702 & 15.888 & 15.300 & 14.693 & 14.320 &   11.28 &  $-$43.18 & 0.805 & memb & UGCS$-$Pl$-$13 \cr
03 44 22.44 & 23 39 01.4 & 19.013 & 17.739 & 16.885 & 16.249 & 15.647 &   28.82 &  $-$42.67 & 0.647 & memb & Roque5     \cr
03 44 23.24 & 25 38 44.9 & 16.288 & 15.349 & 14.692 & 14.133 & 13.744 &   16.08 &  $-$28.17 &  ---  & cand & BRB4       \cr
03 44 25.58 & 22 40 07.9 & 16.194 & 15.511 & 14.944 & 14.383 & 13.989 &   16.23 &  $-$30.26 & 0.658 & memb & UGCS$-$Pl$-$14 \cr
03 44 32.33 & 25 25 18.0 & 16.945 & 16.120 & 15.460 & 14.884 & 14.466 &   20.34 &  $-$37.19 & 0.949 & memb & cfht10     \cr
03 44 34.30 & 23 51 24.6 & 16.799 & 16.016 & 15.444 & 14.858 & 14.453 &   33.10 &  $-$39.85 &  ---  & cand & UGCS$-$Pl$-$15 \cr
03 44 35.16 & 25 13 42.8 & 17.567 & 16.482 & 15.651 & 14.979 & 14.439 &   21.14 &  $-$39.28 & 0.952 & memb & CFHTPLIZ9,CFHT$-$Pl$-$16 \cr
03 44 35.90 & 23 34 42.0 & 16.224 & 15.551 & 14.991 & 14.369 & 13.973 &   21.72 &  $-$32.67 & 0.840 & memb & HHJ5       \cr
03 44 53.12 & 23 34 22.9 & 17.223 & 16.356 & 15.740 & 15.110 & 14.695 &   21.30 &  $-$31.20 & 0.754 & memb & UGCS$-$Pl$-$16 \cr
03 45 04.41 & 24 15 16.6 & 20.325 & 18.895 & 17.768 & 16.943 & 16.318 &    ---  &    ---  &  ---  & cand & UGCS$-$Pl$-$17 \cr
03 45 08.69 & 24 24 09.4 & 16.345 & 15.700 & 15.129 & 14.552 & 14.213 &   18.35 &  $-$52.13 &  ---  & cand & UGCS$-$Pl$-$18 \cr
03 45 09.46 & 23 58 44.7 & 16.820 & 16.099 & 15.417 & 14.830 & 14.382 &   30.39 &  $-$47.73 &  ---  & cand & PPL2       \cr
03 45 31.37 & 24 52 47.5 & 17.246 & 16.226 & 15.471 & 14.842 & 14.329 &   19.65 &  $-$30.23 &  ---  & cand & IPMBD29    \cr
03 45 35.69 & 24 24 34.2 & 16.487 & 15.726 & 15.154 & 14.577 & 14.186 &    9.53 &  $-$47.20 &  ---  & cand & UGCS$-$Pl$-$19 \cr
03 45 37.76 & 23 43 50.1 & 16.129 & 15.320 & 14.710 & 14.149 & 13.731 &   28.02 &  $-$41.71 & 0.696 & memb & UGCS$-$Pl$-$20 \cr
03 45 41.27 & 23 54 09.8 & 17.039 & 16.055 & 15.345 & 14.761 & 14.281 &   15.29 &  $-$42.72 & 0.931 & memb & Roque15    \cr
03 45 42.33 & 24 04 11.2 & 16.420 & 15.757 & 15.203 & 14.624 & 14.252 &   14.24 &  $-$40.62 & 0.931 & memb & UGCS$-$Pl$-$21 \cr
03 45 50.42 & 22 36 05.6 & 17.593 & 16.734 & 16.039 & 15.488 & 15.017 &   20.35 &  $-$39.71 & 0.956 & memb & BPL78      \cr
03 45 50.66 & 24 09 03.5 & 17.352 & 16.451 & 15.692 & 15.060 & 14.569 &   17.66 &  $-$53.57 &  ---  & cand & roque13    \cr
03 45 53.20 & 25 12 55.8 & 17.481 & 16.671 & 15.998 & 15.346 & 14.935 &    ---  &    ---  &  ---  & cand & BPL81      \cr
03 45 54.96 & 23 33 57.9 & 17.032 & 16.209 & 15.560 & 15.005 & 14.550 &   16.40 &  $-$39.61 & 0.952 & memb & UGCS$-$Pl$-$22 \cr
03 46 02.52 & 23 45 33.2 & 18.077 & 17.254 & 16.460 & 15.458 & 15.012 &    ---  &    ---  &  ---  & cand & UGCS$-$Pl$-$23 \cr
03 46 03.75 & 23 44 35.6 & 18.074 & 17.137 & 16.410 & 15.528 & 15.040 &    ---  &    ---  &  ---  & cand & UGCS$-$Pl$-$24 \cr
03 46 08.02 & 23 45 35.5 & 18.787 & 17.829 & 16.857 & 15.832 & 15.324 &    ---  &    ---  &  ---  & cand & UGCS$-$Pl$-$25 \cr
03 46 10.23 & 21 52 55.8 & 16.115 & 15.233 & 14.573 & 13.949 & 13.548 &   36.60 &  $-$41.94 &  ---  & cand & UGCS$-$Pl$-$26 \cr
03 46 14.06 & 23 21 56.5 & 17.005 & 16.220 & 15.574 & 15.013 & 14.589 &   16.37 &  $-$27.55 &  ---  & cand & UGCS$-$Pl$-$27 \cr
03 46 22.25 & 23 52 26.6 & 17.013 & 16.179 & 15.526 & 14.879 & 14.440 &   16.52 &  $-$39.19 & 0.953 & memb & UGCS$-$Pl$-$28 \cr
03 46 23.12 & 24 20 36.1 & 18.081 & 17.027 & 16.233 & 15.636 & 15.134 &    ---  &    ---  &  ---  & cand & BPL100     \cr
03 46 26.09 & 24 05 09.5 & 16.660 & 15.824 & 15.127 & 14.561 & 14.097 &   17.99 &  $-$40.76 & 0.957 & memb & IPMBD25    \cr
03 46 27.10 & 21 48 22.6 & 19.764 & 18.557 & 17.387 & 16.557 & 15.845 &    ---  &    ---  &  ---  & cand & UGCS$-$Pl$-$29 \cr
03 46 32.13 & 24 23 14.6 & 19.104 & 17.932 & 17.030 & 16.345 & 15.840 &    ---  &    ---  &  ---  & cand & UGCS$-$Pl$-$30 \cr
03 46 34.25 & 23 50 03.7 & 19.799 & 18.422 & 17.483 & 16.644 & 16.070 &   17.97 &  $-$38.98 & 0.686 & memb & UGCS$-$Pl$-$31 \cr
03 46 34.99 & 23 31 14.4 & 18.353 & 17.242 & 16.406 & 15.818 & 15.301 &    ---  &    ---  &  ---  & cand & UGCS$-$Pl$-$32 \cr
03 46 35.36 & 23 57 07.4 & 16.768 & 15.943 & 15.346 & 14.756 & 14.384 &   16.65 &  $-$42.51 & 0.944 & memb & UGCS$-$Pl$-$33 \cr
03 46 40.94 & 22 22 38.2 & 19.193 & 18.048 & 16.910 & 16.151 & 15.495 &    ---  &    ---  &  ---  & cand & UGCS$-$Pl$-$34 \cr
03 46 48.56 & 23 09 57.6 & 16.434 & 15.709 & 15.087 & 14.515 & 14.078 &   20.25 &  $-$36.11 &  ---  & cand & UGCS$-$Pl$-$35 \cr
03 46 50.03 & 24 00 23.6 & 17.295 & 16.335 & 15.594 & 14.987 & 14.502 &   18.70 &  $-$36.10 & 0.943 & memb & UGCS$-$Pl$-$36 \cr
03 46 51.82 & 23 23 09.4 & 19.574 & 18.324 & 17.182 & 16.399 & 15.647 &    ---  &    ---  &  ---  & cand & UGCS$-$Pl$-$37 \cr
03 46 52.97 & 24 15 07.8 & 16.271 & 15.623 & 15.058 & 14.480 & 14.125 &   19.58 &  $-$42.79 &  ---  & cand & MHO7       \cr
03 46 55.48 & 23 11 16.1 & 20.298 & 18.965 & 17.886 & 17.117 & 16.426 &    ---  &    ---  &  ---  & cand & UGCS$-$Pl$-$38 \cr
03 47 01.85 & 24 13 28.1 & 16.124 & 15.496 & 14.931 & 14.374 & 14.022 &   18.48 &  $-$43.76 &  ---  & cand & MHO10      \cr
03 47 04.41 & 24 47 27.4 & 20.645 & 19.401 & 18.057 & 17.099 & 16.504 &    ---  &    ---  &  ---  & cand & UGCS$-$Pl$-$39 \cr
03 47 05.71 & 24 40 03.6 & 16.905 & 16.048 & 15.447 & 14.850 & 14.425 &   25.97 &  $-$35.30 &  ---  & cand & BPL124     \cr
03 47 05.79 & 23 45 34.7 & 16.168 & 15.444 & 14.883 & 14.314 & 13.972 &   19.17 &  $-$42.39 & 0.950 & memb & UGCS$-$Pl$-$40 \cr
 \hline
\end{tabular}
\end{table*}

\begin{table*}
  \begin{tabular}{c c c c c c c c c c c l}
  \hline
R.A.\ & Dec.\  &  $Z$  &  $Y$  &  $J$  &  $H$  & $K$ & $\mu_{\alpha}cos\delta$ & $\mu_{\delta}$ & Prob & Memb? & Other name \\
 \hline
03 47 10.65 & 23 58 16.4 & 16.041 & 15.466 & 14.887 & 14.340 & 13.990 &   16.16 &  $-$33.75 & 0.886 & memb & UGCS$-$Pl$-$41 \cr
03 47 11.79 & 24 13 31.3 & 16.182 & 15.444 & 14.805 & 14.234 & 13.860 &    9.97 &  $-$28.34 &  ---  & cand & MHO11      \cr
03 47 17.92 & 24 22 31.7 & 18.033 & 16.955 & 16.192 & 15.581 & 15.093 &    ---  &    ---  &  ---  & cand & Teide1     \cr
03 47 18.10 & 24 45 14.6 & 19.060 & 17.973 & 17.107 & 16.303 & 15.678 &    ---  &    ---  &  ---  & cand & UGCS$-$Pl$-$42 \cr
03 47 27.72 & 22 09 38.6 & 17.332 & 16.426 & 15.733 & 15.169 & 14.726 &    9.37 &  $-$38.49 &  ---  & cand & MHOBD5     \cr
03 47 29.59 & 23 52 49.4 & 16.294 & 15.563 & 15.003 & 14.425 & 14.049 &   14.04 &  $-$37.97 &  ---  & cand & UGCS$-$Pl$-$43 \cr
03 47 39.02 & 24 36 22.2 & 17.068 & 16.158 & 15.537 & 14.954 & 14.530 &   20.06 &  $-$39.23 & 0.956 & memb & BRB12      \cr
03 47 46.77 & 25 35 16.6 & 20.340 & 18.479 & 17.438 & 16.673 & 16.142 &    ---  &    ---  &  ---  & cand & UGCS$-$Pl$-$44 \cr
03 47 48.91 & 24 17 06.6 & 20.153 & 19.148 & 17.841 & 17.014 & 16.406 &    ---  &    ---  &  ---  & cand & UGCS$-$Pl$-$45 \cr
03 47 49.45 & 23 31 52.9 & 17.057 & 16.199 & 15.615 & 15.022 & 14.606 &   34.87 &  $-$44.39 &  ---  & cand & UGCS$-$Pl$-$46 \cr
03 47 50.41 & 23 54 47.9 & 18.087 & 17.007 & 16.300 & 15.589 & 15.078 &    ---  &    ---  &  ---  & cand & NoName     \cr
03 47 58.03 & 22 06 50.9 & 16.686 & 15.892 & 15.282 & 14.721 & 14.327 &   24.86 &  $-$32.10 &  ---  & cand & bpl163     \cr
03 47 59.73 & 22 36 01.9 & 17.981 & 16.986 & 16.206 & 15.588 & 15.106 &    ---  &    ---  &  ---  & cand & int$-$pl$-$IZ$-$33 \cr
03 48 04.67 & 23 39 30.2 & 16.964 & 15.939 & 15.294 & 14.683 & 14.256 &   21.59 &  $-$29.33 &  ---  & cand & PPL15      \cr
03 48 14.30 & 24 15 50.6 & 16.582 & 15.828 & 15.226 & 14.675 & 14.253 &   10.53 &  $-$52.30 &  ---  & cand & BPL169     \cr
03 48 27.36 & 23 46 16.3 & 20.616 & 19.569 & 18.148 & 17.346 & 16.490 &    ---  &    ---  &  ---  & cand & UGCS$-$Pl$-$47 \cr
03 48 30.29 & 24 18 00.3 & 20.176 & 19.108 & 17.804 & 16.960 & 16.373 &    ---  &    ---  &  ---  & cand & UGCS$-$Pl$-$48 \cr
03 48 31.52 & 24 34 37.3 & 19.163 & 17.784 & 16.727 & 15.964 & 15.343 &   18.05 &  $-$39.05 & 0.687 & memb & BRB16      \cr
03 48 38.37 & 22 33 51.8 & 17.262 & 16.416 & 15.707 & 15.162 & 14.713 &    8.64 &  $-$50.82 &  ---  & cand & UGCS$-$Pl$-$49 \cr
03 48 44.69 & 24 37 23.5 & 16.208 & 15.478 & 14.916 & 14.396 & 13.989 &   23.06 &  $-$35.91 & 0.910 & memb & CFHT5      \cr
03 48 55.65 & 24 21 40.2 & 16.160 & 15.535 & 14.953 & 14.391 & 14.031 &   19.28 &  $-$39.16 &  ---  & cand & HHJ8       \cr
03 49 04.86 & 23 33 39.3 & 17.133 & 16.229 & 15.593 & 15.037 & 14.573 &    9.55 &  $-$35.28 &  ---  & cand & Roque47    \cr
03 49 05.18 & 22 04 52.7 & 16.624 & 15.857 & 15.264 & 14.724 & 14.331 &   12.38 &  $-$53.26 &  ---  & cand & UGCS$-$Pl$-$50 \cr
03 49 12.51 & 24 11 12.8 & 18.376 & 17.249 & 16.413 & 15.776 & 15.254 &    ---  &    ---  &  ---  & cand & BPL201     \cr
03 49 15.12 & 24 36 22.5 & 16.797 & 15.947 & 15.367 & 14.846 & 14.418 &   17.46 &  $-$42.66 & 0.946 & memb & BRB10      \cr
03 49 41.21 & 22 56 40.6 & 16.218 & 15.561 & 15.008 & 14.415 & 14.072 &   20.05 &  $-$42.05 & 0.950 & memb & BPL213     \cr
03 49 43.17 & 24 39 46.5 & 16.325 & 15.616 & 15.067 & 14.507 & 14.100 &   18.57 &  $-$34.64 & 0.923 & memb & BPL215     \cr
03 49 52.43 & 24 03 43.0 & 16.028 & 15.423 & 14.871 & 14.330 & 13.947 &   23.88 &  $-$39.12 &  ---  & cand & UGCS$-$Pl$-$51 \cr
03 49 56.81 & 24 59 07.1 & 16.412 & 15.744 & 15.142 & 14.591 & 14.180 &   17.87 &  $-$40.58 & 0.957 & memb & BPL218     \cr
03 50 08.27 & 25 30 51.7 & 19.578 & 18.202 & 17.248 & 16.579 & 16.007 &    ---  &    ---  &  ---  & cand & UGCS$-$Pl$-$52 \cr
03 50 13.39 & 23 59 29.8 & 20.404 & 19.093 & 17.845 & 16.892 & 16.205 &    ---  &    ---  &  ---  & cand & UGCS$-$Pl$-$53 \cr
03 50 16.09 & 24 08 34.8 & 19.872 & 18.458 & 17.338 & 16.664 & 16.077 &   15.01 &  $-$36.98 & 0.619 & memb & Roque30    \cr
03 50 19.15 & 24 16 34.0 & 16.370 & 15.687 & 15.081 & 14.547 & 14.174 &   17.25 &  $-$44.27 & 0.926 & memb & BPL228     \cr
03 50 22.01 & 23 55 30.4 & 17.248 & 16.292 & 15.692 & 15.077 & 14.673 &   29.42 &  $-$25.82 &  ---  & cand & UGCS$-$Pl$-$54 \cr
03 51 05.97 & 24 36 16.9 & 17.082 & 16.231 & 15.654 & 15.122 & 14.703 &   10.72 &  $-$34.15 & 0.664 & memb & CFHT$-$PLIZ1 \cr
03 51 26.60 & 22 48 46.0 & 18.839 & 99.999 & 16.661 & 15.898 & 15.309 &    ---  &    ---  &  ---  & cand & UGCS$-$Pl$-$55 \cr
03 51 38.96 & 24 30 44.8 & 18.723 & 17.316 & 16.408 & 15.702 & 15.148 &   18.71 &  $-$41.66 & 0.751 & memb & UGCS$-$Pl$-$56 \cr
03 51 42.34 & 25 57 25.6 & 16.116 & 15.540 & 14.967 & 14.376 & 14.006 &   24.40 &  $-$26.00 &  ---  & cand & UGCS$-$Pl$-$57 \cr
03 51 44.94 & 23 26 39.3 & 17.732 & 16.805 & 16.031 & 15.401 & 14.971 &   13.26 &  $-$38.62 & 0.912 & memb & CFHTPLIZ10,BPL240 \cr
03 51 59.27 & 23 17 17.8 & 17.351 & 16.490 & 15.811 & 15.245 & 14.859 &   23.58 &  $-$35.81 & 0.898 & memb & UGCS$-$Pl$-$58 \cr
03 52 02.10 & 23 15 45.4 & 18.648 & 17.523 & 16.700 & 16.032 & 15.474 &   20.83 &  $-$37.87 & 0.737 & memb & BPL249     \cr
03 52 05.82 & 24 17 31.1 & 16.470 & 15.771 & 15.186 & 14.614 & 14.251 &   13.32 &  $-$41.59 & 0.909 & memb & NoName     \cr
03 52 06.72 & 24 16 00.5 & 17.040 & 16.168 & 15.523 & 14.967 & 14.503 &   13.37 &  $-$48.12 & 0.638 & memb & CFHT$-$PLIZ3,CFHT$-$Pl$-$13\_Teide2 \cr
03 52 55.92 & 24 57 41.8 & 16.279 & 15.671 & 15.099 & 14.518 & 14.148 &   25.39 &  $-$41.63 & 0.870 & memb & BPL275     \cr
03 53 23.13 & 23 19 20.4 & 17.768 & 16.798 & 15.963 & 15.327 & 14.818 &    ---  &    ---  &  ---  & cand & BPL283     \cr
03 53 24.24 & 25 14 37.7 & 16.722 & 15.951 & 15.332 & 14.772 & 14.381 &   13.82 &  $-$45.98 &  ---  & cand & UGCS$-$Pl$-$59 \cr
03 54 05.35 & 23 33 59.3 & 18.668 & 17.505 & 16.666 & 15.961 & 15.426 &   14.61 &  $-$39.84 & 0.717 & memb & CFHTPLIZ20,CFHT$-$Pl$-$25\_BPL \cr
03 54 15.28 & 25 09 52.2 & 17.828 & 16.913 & 16.164 & 15.550 & 15.071 &   16.71 &  $-$20.27 &  ---  & cand & BPL306     \cr
03 54 31.48 & 22 39 01.6 & 16.552 & 15.766 & 15.173 & 14.571 & 14.187 &   15.83 &  $-$40.39 & 0.948 & memb & UGCS$-$Pl$-$60 \cr
03 55 12.61 & 23 17 37.3 & 17.759 & 16.765 & 15.962 & 15.330 & 14.835 &    ---  &    ---  &  ---  & cand & CFHT15     \cr
03 55 18.11 & 24 17 05.7 & 16.242 & 15.660 & 15.100 & 14.526 & 14.185 &   21.11 &  $-$22.06 &  ---  & cand & BPL326     \cr
03 55 27.06 & 25 14 45.8 & 16.048 & 15.299 & 14.643 & 14.046 & 13.659 &   22.98 &  $-$39.83 & 0.936 & memb & BPL328     \cr
03 55 47.14 & 25 14 39.6 & 17.188 & 16.421 & 15.772 & 15.201 & 14.797 &    8.24 &  $-$29.85 &  ---  & cand & UGCS$-$Pl$-$61 \cr
03 55 47.45 & 22 50 50.2 & 19.883 & 18.435 & 17.421 & 16.667 & 16.088 &    ---  &    ---  &  ---  & cand & UGCS$-$Pl$-$62 \cr
03 56 11.38 & 25 03 36.5 & 16.606 & 15.895 & 15.229 & 14.675 & 14.323 &   22.99 &  $-$41.23 & 0.932 & memb & BPL334     \cr
 \hline
\end{tabular}
\end{table*}
%

%
%
%
%
\section{Table of faint proper motion non-members in the Pleiades
extracted from the UKIDSS GCS DR1\@}

\begin{table*}
  \caption{List of 53 high probability and new photometric candidates
classified as PM non-members based on their PM derived from the
2MASS vs GCS cross-correlation.
This table lists the equatorial coordinates (in J2000),
magnitudes from the GCS, and PMs for each source.
}
  \label{tab_Pleiades:ZJcand_PM_NM}
  \begin{tabular}{c c c c c c c c c}
  \hline
R.A.\ & Dec.\  &  $Z$  &  $Y$  &  $J$  &  $H$  & $K$ & $\mu_{\alpha}cos\delta$ & $\mu_{\delta}$ \\
 \hline
03 41 45.82 & 23 14 26.1 & 16.003 & 15.484 & 14.953 & 14.256 & 13.957 &   14.50 &  $-$14.79 \cr
03 42 04.72 & 23 29 04.2 & 16.183 & 15.652 & 15.107 & 14.438 & 14.118 &   $-$6.28 &  $-$11.03 \cr
03 42 10.17 & 22 48 44.6 & 17.200 & 16.245 & 15.528 & 14.893 & 14.443 &   19.91 &  $-$82.78 \cr
03 43 53.05 & 23 21 50.2 & 16.062 & 15.560 & 15.019 & 14.336 & 14.015 &  $-$23.88 &  $-$14.35 \cr
03 44 09.39 & 23 17 07.2 & 16.319 & 15.751 & 15.168 & 14.586 & 14.233 &  $-$16.32 &    4.32 \cr
03 44 10.08 & 22 43 57.8 & 16.275 & 15.691 & 15.175 & 14.569 & 14.246 &   $-$3.75 &  $-$12.96 \cr
03 44 33.80 & 22 42 49.2 & 16.485 & 15.857 & 15.333 & 14.709 & 14.331 &    2.88 &   18.30 \cr
03 45 21.12 & 21 46 17.5 & 16.262 & 15.793 & 15.149 & 14.562 & 14.230 &  $-$14.43 &  $-$38.91 \cr
03 45 22.14 & 21 52 40.0 & 16.220 & 15.655 & 15.115 & 14.550 & 14.191 &    4.94 &  $-$66.09 \cr
03 45 28.34 & 23 48 09.6 & 16.983 & 16.259 & 15.577 & 14.721 & 14.285 &   21.69 &   $-$0.94 \cr
03 45 29.86 & 22 24 14.6 & 16.335 & 15.699 & 15.176 & 14.555 & 14.221 &   40.98 &  $-$85.07 \cr
03 45 34.50 & 23 41 43.5 & 16.959 & 16.237 & 15.608 & 14.846 & 14.430 &  $-$17.30 &  $-$14.64 \cr
03 45 36.44 & 24 18 15.4 & 16.079 & 15.561 & 15.037 & 14.415 & 14.095 &    8.91 &    6.78 \cr
03 45 36.85 & 23 44 47.8 & 16.669 & 16.245 & 15.423 & 14.641 & 14.244 &    7.10 &   12.74 \cr
03 45 37.25 & 23 49 21.0 & 16.279 & 15.725 & 15.116 & 14.203 & 13.866 &  $-$21.11 &    9.22 \cr
03 45 43.12 & 25 40 23.1 & 16.150 & 14.905 & 13.924 & 13.220 & 12.638 & $-$102.76 &  $-$38.74 \cr
03 45 49.90 & 23 45 59.8 & 16.124 & 15.484 & 14.810 & 13.887 & 13.578 &    6.17 &    7.13 \cr
03 46 08.22 & 23 21 38.7 & 17.011 & 16.367 & 15.720 & 14.986 & 14.614 &  $-$14.77 &   $-$6.54 \cr
03 46 14.47 & 22 20 51.1 & 16.639 & 15.992 & 15.456 & 14.964 & 14.587 &   72.99 &  $-$67.54 \cr
03 46 26.76 & 24 49 18.1 & 16.210 & 15.604 & 15.109 & 14.497 & 14.196 &  $-$30.58 &  $-$28.37 \cr
03 46 33.00 & 23 38 00.9 & 16.537 & 15.946 & 15.377 & 14.425 & 14.120 &   $-$9.39 &    7.18 \cr
03 46 34.16 & 23 25 12.5 & 16.210 & 15.672 & 15.079 & 14.377 & 14.041 &    9.28 &    4.89 \cr
03 46 36.83 & 23 33 01.8 & 16.659 & 15.959 & 15.359 & 14.488 & 14.113 &   $-$6.73 &  $-$11.83 \cr
03 46 44.04 & 23 38 13.5 & 17.114 & 16.411 & 15.786 & 14.916 & 14.564 &   20.68 &   $-$7.25 \cr
03 46 50.99 & 25 40 44.6 & 16.230 & 15.655 & 15.164 & 14.450 & 14.174 &   $-$2.52 &  $-$13.87 \cr
03 46 52.29 & 21 47 43.1 & 16.695 & 15.996 & 15.417 & 14.783 & 14.401 &   22.89 &   $-$8.81 \cr
03 47 06.65 & 24 45 47.4 & 16.051 & 15.414 & 14.968 & 14.414 & 14.093 &   75.15 &   40.22 \cr
03 47 08.31 & 22 33 10.0 & 16.500 & 15.881 & 15.332 & 14.804 & 14.422 &   $-$9.26 &   17.59 \cr
03 47 13.69 & 23 46 28.4 & 16.347 & 15.706 & 15.221 & 14.656 & 14.329 &  $-$14.07 &    7.48 \cr
03 47 36.27 & 24 28 50.1 & 16.049 & 15.482 & 14.967 & 14.279 & 13.929 &   $-$9.18 &   $-$5.59 \cr
03 48 23.62 & 24 22 35.2 & 16.151 & 15.466 & 14.890 & 14.299 & 13.938 &  $-$15.18 &   $-$6.70 \cr
03 48 36.30 & 23 33 25.3 & 16.054 & 15.512 & 15.035 & 14.472 & 14.147 &  $-$16.67 &  $-$28.01 \cr
03 48 48.46 & 21 59 00.3 & 16.514 & 15.869 & 15.332 & 14.779 & 14.470 &  $-$18.96 &    3.15 \cr
03 49 11.59 & 24 26 17.5 & 16.030 & 15.415 & 14.930 & 14.412 & 14.078 &   35.55 &  $-$81.10 \cr
03 49 12.12 & 23 12 55.9 & 16.821 & 16.077 & 15.427 & 14.815 & 14.414 &   67.42 &    4.88 \cr
03 50 03.94 & 24 56 02.9 & 16.179 & 15.630 & 15.102 & 14.494 & 14.124 &   $-$6.50 &  $-$42.38 \cr
03 50 15.55 & 26 06 30.1 & 16.821 & 16.081 & 15.407 & 14.810 & 14.390 &   45.38 &  $-$60.60 \cr
03 51 04.04 & 24 32 57.9 & 16.032 & 15.447 & 14.981 & 14.371 & 14.053 &   83.02 &  $-$39.72 \cr
03 51 09.72 & 25 18 52.4 & 16.705 & 15.956 & 15.399 & 14.872 & 14.515 &   $-$8.24 &  $-$31.78 \cr
03 51 10.52 & 22 48 14.5 & 16.717 & 16.031 & 15.451 & 14.754 & 14.406 &   $-$9.03 &  $-$22.58 \cr
03 51 27.90 & 22 48 12.9 & 16.073 & 15.332 & 14.720 & 14.081 & 13.693 &   24.54 &    5.16 \cr
03 51 38.18 & 23 03 11.2 & 16.859 & 16.116 & 15.533 & 14.952 & 14.580 &   $-$8.79 &   $-$0.54 \cr
03 52 07.88 & 23 59 13.1 & 16.135 & 15.284 & 14.620 & 14.057 & 13.620 &   22.45 &  $-$86.80 \cr
03 52 17.49 & 22 51 01.9 & 16.583 & 15.973 & 15.412 & 14.759 & 14.389 &   $-$8.79 &    3.35 \cr
03 53 48.72 & 25 04 20.1 & 16.729 & 16.213 & 15.240 & 15.098 & 14.797 &    3.44 &  $-$18.98 \cr
03 53 59.92 & 23 41 00.3 & 16.175 & 15.572 & 15.052 & 14.450 & 14.115 &   $-$5.34 &  $-$13.33 \cr
03 54 17.46 & 23 11 56.9 & 16.131 & 15.574 & 15.017 & 14.432 & 14.088 &   $-$9.91 &    5.36 \cr
03 54 39.33 & 23 03 12.4 & 16.394 & 15.588 & 14.992 & 14.444 & 14.053 &   10.21 &   $-$1.55 \cr
03 54 40.56 & 23 42 23.7 & 16.106 & 15.547 & 15.060 & 14.425 & 14.155 &    7.21 &    4.10 \cr
03 54 46.11 & 23 00 20.6 & 17.017 & 16.309 & 15.704 & 15.103 & 14.741 &  $-$26.85 &   $-$1.64 \cr
03 54 54.51 & 22 50 21.5 & 16.403 & 15.752 & 15.216 & 14.651 & 14.338 &   $-$0.85 &   $-$0.54 \cr
03 55 14.72 & 22 42 08.5 & 16.637 & 15.962 & 15.384 & 14.731 & 14.381 &   23.51 &  $-$12.75 \cr
03 55 30.07 & 23 54 53.5 & 16.466 & 15.829 & 15.296 & 14.788 & 14.421 &   $-$7.60 &   $-$1.05 \cr
 \hline
\end{tabular}
\end{table*}
%

%
%
%
%
\section{Table of faint photometric non-members in the Pleiades
extracted from the UKIDSS GCS DR1\@}

\begin{table*}
  \caption{List of 24 high probability and new photometric candidates
classified as photometric non-members from their location in several
colour-magnitude diagrams.
This table lists the equatorial coordinates (in J2000),
magnitudes from the GCS, and PMs for each source.
}
  \label{tab_Pleiades:ZJcand_photNM}
  \begin{tabular}{c c c c c c c c c}
  \hline
R.A.\ & Dec.\  &  $Z$  &  $Y$  &  $J$  &  $H$  & $K$ & $\mu_{\alpha}cos\delta$ & $\mu_{\delta}$ \\
 \hline
03 41 31.46 & 23 05 12.3 & 18.520 & 19.416 & 16.757 & 16.269 & 16.179 &    ---  &    ---  \cr
03 41 36.55 & 22 41 01.7 & 16.248 & 15.641 & 15.136 & 14.585 & 14.250 &   35.42 &  $-$33.96 \cr
03 42 27.94 & 25 25 20.0 & 19.822 & 19.332 & 17.711 & 17.062 & 16.874 &    ---  &    ---  \cr
03 43 12.21 & 23 09 16.6 & 18.540 & 17.489 & 16.670 & 16.022 & 15.589 &    ---  &    ---  \cr
03 43 52.03 & 22 55 24.5 & 18.874 & 17.727 & 16.970 & 16.272 & 15.782 &    ---  &    ---  \cr
03 44 12.68 & 25 24 35.1 & 18.058 & 17.144 & 16.455 & 15.949 & 15.527 &    ---  &    ---  \cr
03 45 33.16 & 25 34 30.0 & 18.079 & 17.112 & 16.422 & 15.843 & 15.381 &    ---  &    ---  \cr
03 46 04.92 & 22 15 40.2 & 17.386 & 16.555 & 15.899 & 15.246 & 15.018 &    ---  &    ---  \cr
03 47 46.15 & 25 21 42.3 & 19.444 & 18.280 & 17.382 & 16.876 & 16.331 &    ---  &    ---  \cr
03 48 19.02 & 24 25 12.8 & 17.666 & 16.597 & 15.972 & 15.375 & 14.951 &   19.35 &  $-$34.91 \cr
03 48 49.03 & 24 20 25.3 & 19.165 & 18.125 & 17.229 & 16.543 & 16.037 &    ---  &    ---  \cr
03 48 58.61 & 23 37 03.9 & 19.567 & 18.221 & 17.389 & 16.735 & 16.231 &    ---  &    ---  \cr
03 49 21.17 & 23 34 02.0 & 18.409 & 17.353 & 16.637 & 16.030 & 15.530 &    ---  &    ---  \cr
03 49 48.77 & 23 42 59.4 & 19.147 & 18.083 & 17.235 & 16.608 & 16.099 &    ---  &    ---  \cr
03 49 51.23 & 25 26 06.6 & 19.698 & 18.476 & 17.608 & 17.059 & 16.494 &    ---  &    ---  \cr
03 50 15.33 & 24 35 40.4 & 16.124 & 15.521 & 15.070 & 14.470 & 14.175 &    8.32 &  $-$22.94 \cr
03 51 37.72 & 25 42 46.6 & 19.070 & 17.672 & 16.338 & 15.428 & 14.640 &    ---  &    ---  \cr
03 51 59.93 & 23 24 25.6 & 19.627 & 18.336 & 17.387 & 16.694 & 16.250 &    ---  &    ---  \cr
03 52 13.44 & 24 28 52.3 & 20.026 & 18.527 & 17.772 & 17.224 & 16.643 &    ---  &    ---  \cr
03 52 46.44 & 24 24 16.9 & 19.536 & 18.447 & 17.533 & 16.955 & 16.317 &    ---  &    ---  \cr
03 53 26.55 & 24 46 44.9 & 16.022 & 15.487 & 14.996 & 14.417 & 14.107 &   $-$2.92 &  $-$47.58 \cr
03 54 55.92 & 24 37 43.0 & 20.857 & 19.356 & 18.324 & 17.897 & 17.734 &    ---  &    ---  \cr
03 55 39.58 & 24 12 51.1 & 20.442 & 19.165 & 17.853 & 17.285 & 16.634 &    ---  &    ---  \cr
03 55 42.01 & 22 57 01.4 & 18.624 & 17.246 & 15.954 & 14.995 & 14.189 &    ---  &    ---  \cr
 \hline
\end{tabular}
\end{table*}
%

%
%
%
%
\section{Table of faint $YJ$ candidates in the Pleiades
extracted from the UKIDSS GCS DR1\@.}

\begin{table*}
  \caption{Near-infrared ($ZYJHK$) photometry for 35
photometric candidates with no $Z$ detection and
selected from the ($Y-J$,$Y$)
CMD\@. This table lists the equatorial
coordinates (in J2000), the magnitudes from the GCS,
their PMs, and their membership status
(Memb$\equiv$photometric and PM member;
cand$\equiv$photometric candidate
in all diagrams; photNM$\equiv$photometric non-member;
PM\_NM$\equiv$ PM non-member; dubious$\equiv$likely false
detection).
}
  \label{tab_Pleiades:YJcand}
  \begin{tabular}{c c c c c c c c l}
  \hline
R.A.\ & Dec.\  &  $Y$  &  $J$  &  $H$  & $K$ & $\mu_{\alpha}cos\delta$ & $\mu_{\delta}$ & Memb? \\ \hline
03 42 14.28 & 22 43 02.5 & 19.977 & 18.526 & 18.017 & 17.482 &     ---  &     ---  & photNM  \cr
03 42 59.30 & 25 37 39.1 & 19.727 & 18.241 & 17.367 & 16.647 &     ---  &     ---  & Memb    \cr
03 44 30.52 & 24 21 17.4 & 19.458 & 18.396 & 17.684 & 17.301 &     ---  &     ---  & dubious \cr
03 44 31.28 & 25 35 14.8 & 19.336 & 18.282 & 17.394 & 16.658 &    13.96 & $-$39.80 & Memb    \cr
03 44 47.32 & 24 21 35.8 & 19.553 & 18.411 & 17.477 & 16.898 &     ---  &     ---  & photNM  \cr
03 45 01.81 & 24 04 17.8 & 20.431 & 19.025 & 18.800 & 18.030 &     ---  &     ---  & dubious \cr
03 45 11.56 & 23 25 35.1 & 20.233 & 18.911 & 18.104 & 17.485 &     ---  &     ---  & photNM  \cr
03 45 33.30 & 23 34 34.3 & 19.576 & 18.130 & 17.182 & 16.485 &    13.67 &  $-$9.64 & PM\_NM  \cr
03 45 35.06 & 21 56 22.3 & 19.364 & 18.168 & 17.575 & 17.114 & $-$11.06 &     1.62 & PM\_NM  \cr
03 46 27.94 & 23 42 38.9 & 19.681 & 18.570 & 17.708 & 17.437 &  $-$8.48 & $-$26.82 & PM\_NM  \cr
03 46 29.11 & 22 59 47.7 & 18.992 & 17.740 & 16.768 & 15.924 &     ---  &     ---  & Memb    \cr
03 46 51.05 & 22 34 28.9 & 19.779 & 18.581 & 18.075 & 17.292 &     ---  &     ---  & dubious \cr
03 47 33.15 & 23 36 32.6 & 19.210 & 17.894 & 18.063 & 16.898 &     ---  &     ---  & dubious \cr
03 47 38.47 & 23 56 27.7 & 19.827 & 18.334 & 17.435 & 16.661 &     ---  &     ---  & Memb    \cr
03 47 44.16 & 24 57 24.1 & 20.088 & 18.692 & 17.877 & 17.246 &     ---  &     ---  & photNM  \cr
03 47 46.52 & 24 55 46.6 & 19.426 & 18.247 & 17.398 & 16.590 &     ---  &     ---  & Memb    \cr
03 48 05.47 & 21 57 31.2 & 20.043 & 18.753 & 17.929 & 17.311 &     ---  &     ---  & photNM  \cr
03 48 15.64 & 25 50 09.0 & 19.785 & 18.497 & 17.631 & 16.758 &    25.82 &  $-$44.45 & Memb    \cr
03 48 39.56 & 23 56 11.7 & 20.619 & 19.223 & 18.728 & 18.153 &     ---  &     ---  & photNM  \cr
03 49 17.16 & 23 19 42.8 & 19.726 & 18.548 & 17.657 & 16.738 &     ---  &     ---  & Memb    \cr
03 49 28.92 & 23 22 49.1 & 19.975 & 18.385 & 18.095 & 16.706 &     ---  &     ---  & dubious \cr
03 50 15.96 & 24 23 28.7 & 20.017 & 18.711 & 17.774 & 16.887 &     ---  &     ---  & Memb    \cr
03 50 39.54 & 25 02 54.7 & 19.677 & 18.208 & 17.331 & 16.565 &     ---  &     ---  & Memb    \cr
03 51 29.47 & 24 00 37.4 & 19.610 & 18.412 & 17.463 & 16.696 &    25.57 &  $-$44.48 & Memb    \cr
03 51 41.62 & 25 55 45.4 & 20.484 & 19.107 & 18.453 & 18.047 &     ---  &     ---  & photNM  \cr
03 52 05.33 & 25 37 34.0 & 18.874 & 17.699 & 17.937 & 17.520 &     ---  &     ---  & dubious \cr
03 52 27.18 & 23 12 08.1 & 19.216 & 18.005 & 17.069 & 16.351 &     1.58 &  $-$9.79 & PM\_NM  \cr
03 52 34.75 & 22 56 04.5 & 19.524 & 18.437 & 17.554 & 17.070 &     ---  &     ---  & photNM  \cr
03 52 39.15 & 24 46 29.5 & 19.189 & 18.069 & 17.098 & 16.486 &  $-$1.52 & $-$33.36 & Memb    \cr
03 52 59.62 & 24 42 35.6 & 20.584 & 19.093 & 18.681 & 18.220 &     ---  &     ---  & dubious \cr
03 53 18.93 & 23 12 39.1 & 19.959 & 18.327 & 17.604 & 16.880 &     ---  &     ---  & photNM  \cr
03 54 10.28 & 23 41 40.1 & 19.156 & 18.124 & 17.141 & 16.377 &     7.50 & $-$27.69 & Memb    \cr
03 54 30.49 & 25 11 21.8 & 19.884 & 18.657 & 18.116 & 17.519 &     ---  &     ---  & photNM  \cr
03 54 49.89 & 24 16 23.4 & 19.451 & 18.279 & 18.241 & 17.260 &     ---  &     ---  & dubious \cr
03 55 08.18 & 23 58 08.7 & 19.528 & 18.388 & 17.863 & 17.393 &     ---  &     ---  & dubious \cr
 \hline
\end{tabular}
\end{table*}
%

%
%
%
%
\section{Table of faint $JK$ candidates in the Pleiades
extracted from the UKIDSS GCS DR1\@.}

\begin{table*}
  \caption{Near-infrared ($ZYJHK$) photometry for 16
candidates selected from the ($J-K$,$J$) CMD and recovered
in the optical surveys (INT$+$CFHT).
This table lists the equatorial
coordinates (in J2000), infrared magnitudes,
PMs, and membership status of each object.
Only one source is likely to be a Pleiades member,
two are classified as photometric non-member (PM\_NM),
and the remaining are likely to be false detection
(``dubious'') after examination of the finding charts 
(note that none lie within a circle of radius
25 mas/yr centered on the Pleiades mean motion, making
their membership highly improbable).
}
  \label{tab_Pleiades:JKcand}
  \begin{tabular}{c c c c c c c l}
  \hline
R.A.\ & Dec.\  &  $J$  &  $H$  & $K$ & $\mu_{\alpha}cos\delta$ & $\mu_{\delta}$ & Memb? \\ \hline
03 50 41.16 & 25 44 24.2 & 18.776 & 17.580 & 16.556 &  $-$15.09 &  $-$16.61 & dubious \cr
03 48 32.69 & 25 06 05.1 & 18.539 & 17.993 & 17.412 &     15.82 &  $-$12.52 & dubious \cr
03 55 57.94 & 24 41 41.6 & 18.876 & 18.173 & 18.215 &   $-$2.06 &      0.57 & dubious \cr
03 55 15.80 & 24 49 32.6 & 17.727 & 17.004 & 16.698 &   $-$8.77 &      8.36 & dubious \cr
03 51 35.09 & 24 03 36.9 & 18.247 & 17.319 & 16.492 &   $-$9.06 &  $-$25.29 & dubious \cr
03 54 47.20 & 23 56 48.0 & 18.845 & 18.429 & 18.556 &  $-$19.94 &     10.38 & dubious \cr
03 55 45.48 & 23 51 25.5 & 18.855 & 18.416 & 18.242 &      2.55 &      6.91 & dubious \cr
03 54 13.41 & 23 32 22.2 & 18.674 & 18.123 & 18.104 &  $-$23.27 &   $-$2.48 & dubious \cr
03 45 35.26 & 23 36 39.6 & 18.898 & 18.388 & 18.123 &     17.73 &   $-$5.33 & dubious \cr
03 44 19.50 & 22 39 03.5 & 18.718 & 18.342 & 17.709 &  $-$30.75 &  $-$16.71 & dubious \cr
03 42 43.46 & 22 38 31.5 & 18.301 & 17.735 & 17.398 &   $-$4.35 &      0.25 & dubious \cr
03 43 24.88 & 22 50 22.2 & 18.018 & 17.310 & 17.065 &   $-$0.25 &      1.76 & dubious \cr
03 52 47.30 & 22 38 42.4 & 17.579 & 16.773 & 15.969 &  $-$13.24 &      0.48 & dubious \cr
03 52 54.90 & 24 37 18.2 & 18.798 & 17.742 & 16.922 &     14.70 &  $-$44.77 & Memb    \cr
03 40 30.34 & 25 58 26.0 & 18.856 & 18.271 & 17.747 &  $-$28.78 &   $-$5.64 & PM\_NM   \cr
03 47 51.19 & 25 26 57.9 & 18.200 & 17.745 & 17.611 &  $-$28.59 &   $-$6.22 & PM\_NM   \cr
 \hline
\end{tabular}
\end{table*}
%

%
%
%
%
\section{Table of low-mass multiple system candidates in the 
Pleiades extracted from the UKIDSS GCS DR1\@.}

\begin{table*}
  \caption{Near-infrared ($ZYJHK$) photometry for 36
photometric low-mass multiple system candidates below
$K$ = 13.5 mag, corresponding to masses of 0.135 M$_{\odot}$
in the Pleiades. This table lists the equatorial
coordinates (in J2000), the magnitudes from the GCS,
and their PMs. The estimated mass of the primary
and the secondary along with the mass ratio (q) are also
given for a total mass less than 70 M$_{\rm Jup}$.
}
  \label{tab_Pleiades:list_binary}
  \begin{tabular}{c c c c c c c c c c c}
  \hline
R.A.\ & Dec.\  &  $Z$  &  $Y$  &  $J$  &  $H$  & $K$ & $\mu_{\alpha}cos\delta$ & $\mu_{\delta}$ & Mass  & q \\ \hline
03 41 40.90 & 25 54 24.1 & 16.829 & 15.900 & 15.173 & 14.562 & 14.114 &   23.78 &  $-$37.48 & 60+60  & 1.00 \cr
03 41 54.16 & 23 05 04.8 & 17.336 & 16.297 & 15.499 & 14.912 & 14.385 &   10.66 &  $-$24.90 & 55+55  & 1.00 \cr
03 42 07.99 & 22 39 33.5 & 18.717 & 17.397 & 16.515 & 15.809 & 15.191 &    ---  &      ---  & 40+35  & 0.87 \cr
03 42 59.92 & 22 42 51.5 & 16.067 & 15.267 & 14.666 & 14.109 & 13.708 &   25.70 &  $-$41.46 &  ---   & 0.00 \cr
03 43 34.48 & 25 57 30.6 & 16.471 & 15.605 & 14.888 & 14.322 & 13.896 &    9.23 &  $-$40.49 & 70+70  & 1.00 \cr
03 44 22.14 & 23 10 54.8 & 15.753 & 15.130 & 14.495 & 13.890 & 13.503 &   19.04 &  $-$39.79 &  ---   & 0.00 \cr
03 44 23.24 & 25 38 44.9 & 16.288 & 15.349 & 14.692 & 14.133 & 13.744 &   16.08 &  $-$28.17 &  ---   & 0.00 \cr
03 44 35.16 & 25 13 42.8 & 17.567 & 16.482 & 15.651 & 14.979 & 14.439 &   21.14 &  $-$39.28 & 50+50  & 1.00 \cr
03 44 35.90 & 23 34 42.0 & 16.224 & 15.551 & 14.991 & 14.369 & 13.973 &   21.72 &  $-$32.67 &  ---   & 0.00 \cr
03 45 09.46 & 23 58 44.7 & 16.820 & 16.099 & 15.417 & 14.830 & 14.382 &   30.39 &  $-$47.73 & 65+40  & 0.62 \cr
03 45 31.37 & 24 52 47.5 & 17.246 & 16.226 & 15.471 & 14.842 & 14.329 &   19.65 &  $-$30.23 & 55+55  & 1.00 \cr
03 45 37.76 & 23 43 50.1 & 16.129 & 15.320 & 14.710 & 14.149 & 13.731 &   28.02 &  $-$41.71 &  ---   & 0.00 \cr
03 45 41.27 & 23 54 09.8 & 17.039 & 16.055 & 15.345 & 14.761 & 14.281 &   15.29 &  $-$42.72 & 70+40  & 0.57 \cr
03 45 50.66 & 24 09 03.5 & 17.352 & 16.451 & 15.692 & 15.060 & 14.569 &   17.66 &  $-$53.57 & 60+35  & 0.58 \cr
03 46 02.52 & 23 45 33.2 & 18.077 & 17.254 & 16.460 & 15.458 & 15.012 &    ---  &      ---  & 40+40  & 1.00 \cr
03 46 03.75 & 23 44 35.6 & 18.074 & 17.137 & 16.410 & 15.528 & 15.040 &    ---  &      ---  & 40+40  & 1.00 \cr
03 46 08.02 & 23 45 35.5 & 18.787 & 17.829 & 16.857 & 15.832 & 15.324 &    ---  &      ---  & 35+35  & 1.00 \cr
03 46 10.23 & 21 52 55.8 & 16.114 & 15.233 & 14.572 & 13.948 & 13.547 &   36.60 &  $-$41.94 &  ---   & 0.00 \cr
03 46 22.25 & 23 52 26.6 & 17.013 & 16.179 & 15.526 & 14.879 & 14.440 &   16.52 &  $-$39.19 & 65+40  & 0.65 \cr
03 46 26.09 & 24 05 09.5 & 16.660 & 15.824 & 15.127 & 14.561 & 14.097 &   17.99 &  $-$40.76 & 60+60  & 1.00 \cr
03 46 27.10 & 21 48 22.6 & 19.764 & 18.557 & 17.387 & 16.557 & 15.845 &    ---  &      ---  & 30+30  & 1.00 \cr
03 46 29.11 & 22 59 47.7 & 99.999 & 18.992 & 17.740 & 16.768 & 15.924 &    ---  &      ---  &  ---   & 0.00 \cr
03 46 40.94 & 22 22 38.2 & 19.193 & 18.048 & 16.910 & 16.151 & 15.495 &    ---  &      ---  & 35+30  & 0.86 \cr
03 46 48.56 & 23 09 57.6 & 16.434 & 15.709 & 15.087 & 14.515 & 14.078 &   20.25 &  $-$36.11 &  ---   & 0.00 \cr
03 46 50.03 & 24 00 23.6 & 17.295 & 16.335 & 15.594 & 14.987 & 14.502 &   18.70 &  $-$36.10 & 60+40  & 0.62 \cr
03 46 51.82 & 23 23 09.4 & 19.574 & 18.324 & 17.182 & 16.399 & 15.647 &    ---  &      ---  & 30+30  & 1.00 \cr
03 47 02.35 & 23 32 36.0 & 15.568 & 14.861 & 14.294 & 13.715 & 13.320 &   20.44 &  $-$42.56 &  ---   & 0.00 \cr
03 47 11.79 & 24 13 31.3 & 16.182 & 15.444 & 14.805 & 14.234 & 13.860 &    9.97 &  $-$28.34 &  ---   & 0.00 \cr
03 48 04.67 & 23 39 30.2 & 16.964 & 15.939 & 15.294 & 14.683 & 14.256 &   21.59 &  $-$29.33 & 65+50  & 0.77 \cr
03 48 31.52 & 24 34 37.3 & 19.163 & 17.784 & 16.727 & 15.964 & 15.343 &   18.05 &  $-$39.05 & 35+35  & 1.00 \cr
03 50 13.39 & 23 59 29.8 & 20.404 & 19.093 & 17.845 & 16.892 & 16.205 &    ---  &      ---  & 30+20  & 0.67 \cr
03 51 38.96 & 24 30 44.8 & 18.723 & 17.316 & 16.408 & 15.702 & 15.148 &   18.71 &  $-$41.66 & 40+35  & 0.87 \cr
03 52 51.79 & 23 33 48.0 & 15.881 & 15.353 & 14.807 & 14.114 & 13.761 &   21.92 &  $-$43.64 &  ---   & 0.00 \cr
03 53 23.13 & 23 19 20.4 & 17.768 & 16.798 & 15.963 & 15.327 & 14.818 &    ---  &      ---  & 50+35  & 0.70 \cr
03 55 12.61 & 23 17 37.3 & 17.758 & 16.765 & 15.962 & 15.330 & 14.835 &    ---  &      ---  & 50+35  & 0.70 \cr
03 55 27.06 & 25 14 45.8 & 16.048 & 15.299 & 14.643 & 14.046 & 13.659 &   22.98 &  $-$39.83 &  ---   & 0.00 \cr
 \hline
\end{tabular}
\end{table*}

\label{lastpage}

\end{document}